\documentclass[10pt,aps,pra,onecolumn,floatfix,superscriptaddress,nofootinbib,longbibliography]{revtex4-2}

\makeatletter
\@ifl@t@r\fmtversion{2023-06-01}{%
  \def\label#1{\@bsphack
    \begingroup
    \UseHookWithArguments{label}{1}{#1}%
    \protected@write\@auxout{}%
           {\string\newlabel{#1}{{\@currentlabel}{\thepage}%
            {\@currentlabelname}{\@currentHref}{\@kernel@reserved@label@data}}}%
    \endgroup
    \@esphack}%
}{}%
\let\REVTeX@orig@class@info\class@info
\def\class@info#1{%
  \begingroup
  \def\REVTeX@msg{Unrecognized LaTeX tabular. Please update this document class! (Proceeding with fingers crossed.)}%
  \def\REVTeX@arg{#1}%
  \ifx\REVTeX@arg\REVTeX@msg
    \endgroup
  \else
    \endgroup\REVTeX@orig@class@info{#1}%
  \fi
}%
\makeatother

\usepackage[T1]{fontenc}
\usepackage[utf8]{inputenc}
\usepackage{amsmath,amssymb,amsfonts,bm}
\usepackage{mathtools}
\usepackage{graphicx}
\graphicspath{{figs/}{./}}
\DeclareGraphicsExtensions{.pdf,.png,.jpg,.jpeg}
\usepackage{xcolor}
\usepackage[colorlinks=true,linkcolor=blue,citecolor=blue,urlcolor=blue]{hyperref}

\newcommand{\D}{\mathcal D}

\newcommand{\Rres}{\mathcal R}
\newcommand{\E}{\mathbb E}

\newcommand{\ii}{\mathrm i}
\newcommand{\ee}{\mathrm e}
\newcommand{\Tr}{\mathrm{Tr}}
\newcommand{\sgn}{\operatorname{sgn}}

\newcommand{\eff}{\mathrm{eff}}
\newcommand{\peak}{\mathrm{peak}}

\newcommand{\eps}{\varepsilon}

\begin{document}

\title{\texorpdfstring{Spectral theory of energy-selective quantum search with Ising Hamiltonian phase oracles}{Spectral theory of energy-selective quantum search with Ising Hamiltonian phase oracles}}
\thanks{Accepted for publication in Physical Review A. DOI: \href{https://doi.org/10.1103/tgsx-5nz4}{10.1103/tgsx-5nz4}. \copyright\ 2026 American Physical Society.}

\author{A. S. Plyashechnik}
\affiliation{Dukhov Research Institute of Automatics (VNIIA), Moscow 127030, Russia}
\author{A. A. Zhukov}
\affiliation{Dukhov Research Institute of Automatics (VNIIA), Moscow 127030, Russia}
\author{A. V. Lebedev}
\affiliation{Dukhov Research Institute of Automatics (VNIIA), Moscow 127030, Russia}
\affiliation{Moscow Institute of Physics and Technology, Dolgoprudny, 141700, Russia}
\author{W. V. Pogosov}
\affiliation{Dukhov Research Institute of Automatics (VNIIA), Moscow 127030, Russia}
\affiliation{Moscow Institute of Physics and Technology, Dolgoprudny, 141700, Russia}
\affiliation{Institute for Theoretical and Applied Electrodynamics, Russian Academy of Sciences, Moscow 125412, Russia}

\begin{abstract}
We develop an exact spectral-response theory for the Grover-type iterate \(W_T=D_\xi\exp(-\ii T H)\), in which the evolution generated by a diagonal Ising Hamiltonian is used directly as a continuous phase oracle. An energy-grouped recurrence and its generating-function solution show how the empirical characteristic function determines the position, width, height, and saturation time of an energy-selective resonance. For an annealed Gaussian density of states, a high-density-tail resonance containing \(M\) configurations is reached after \(\Theta(\sqrt{2^n/M})\) oracle calls with success probability \(\Theta(1)\), giving a quadratic query improvement over independent uniform sampling with classical energy evaluation. For correlated random Ising spectra, overlap-dependent covariances lead to a realization-dependent resonance shift with root-mean-square scale \(O(n^{-3})\), parametrically larger than the resonance width, and can also reduce the peak height. The shift is both an algorithmic detuning and a coherent probe of sample-specific spectral fluctuations whose ensemble statistics reflect the Ising overlap structure. Spectral symmetrization and iterative calibration can remove or compensate the resonance-center displacement for prescribed-energy targeting. We also clarify the relation to designed spectral filters and the precision and coherence requirements of this asymptotic primitive.
\end{abstract}

\maketitle

\section{Introduction}
\label{sec:introduction}

Ising Hamiltonians occupy a special position at the interface of statistical
physics, optimization, and quantum information. They describe spin glasses and
disordered magnets, with the Sherrington--Kirkpatrick model, the random-energy
model, and random-exchange Ising systems serving as standard reference points
for correlated and uncorrelated disorder \cite{SK,Derrida1981,Fan1969}. At the
same time, many NP-hard optimization problems can be encoded as Ising or QUBO
Hamiltonians \cite{Lucas2014}. This dual role explains why Ising models appear
as problem Hamiltonians in quantum annealing and adiabatic computation
\cite{KadowakiNishimori1998,AnnealingFarhi2001,AlbashLidar2018}, in coherent
and other analog Ising machines
\cite{Mohseni2022,Inagaki2016,McMahon2016,Yamamoto2017,Honjo2021,Rah2026}, and
in gate-model approaches such as QAOA, quantum alternating-operator circuits,
Grover-mixer constructions, and variational quantum algorithms
\cite{QAOA,Hadfield2019,Cerezo2021,Tilly2022,Kiktenko2025}. In all these
settings, performance is strongly affected by the structure of the Ising energy
landscape, embedding overheads, and the phase-space organization of the
dynamics \cite{Leleu2019,Hamerly2019,Dobrynin2024,Zhou2025}.

Quantum search provides a complementary way to view such landscapes. In its
standard form, Grover search and amplitude amplification use a Boolean oracle to
find a marked set with a quadratic reduction in the number of oracle calls
\cite{Grover1996,Bennett1997,Boyer1998,Brassard2002,Nielsen2010}, and quantum
minimum finding builds on this idea for optimization \cite{DurrHoyer1996}. For
an Ising objective, however, a conventional Boolean phase oracle would have to
be engineered explicitly: the Ising energy of a basis configuration would be
computed into ancillary qubits, compared with a threshold or target energy, used
to apply a conditional phase flip, and then erased by running the energy
computation backwards. Grover Adaptive Search is a canonical realization of
this comparator-based route for polynomial binary optimization; its oracle uses
reversible arithmetic and ancillary registers to encode and compare objective
values \cite{Gilliam2021}. The Hamiltonian-oracle approach considered here bypasses
this compute--compare--erase construction. A device that implements the
diagonal Ising evolution directly assigns an energy-dependent phase,
\begin{equation}
    U_T=\ee^{-\ii T H},
    \qquad
    U_T|s\rangle=\ee^{-\ii T E_s}|s\rangle .
\end{equation}
Thus \(U_T\) acts as a spectral phase oracle: it converts energy into phase
without first constructing an explicit Boolean marking rule.

The closest comparison is with designed spectral filters. Quantum signal
processing and quantum singular-value transformation (QSP/QSVT) approximate
spectral projectors through polynomial transformations of block-encoded
Hamiltonians \cite{LinTong361,LinTong372}. Quantum eigenvalue transformation of
unitary matrices (QET-U) instead uses Hamiltonian time evolution, normally in
controlled form with one ancilla; for special spin Hamiltonians the control can
be removed \cite{Dong2022}. Fourier/linear-combination-of-unitaries (LCU) and
Chebyshev constructions provide other projector approximations
\cite{GeTuraCirac}, while quantum phase
estimation resolves energy in an auxiliary register before interval selection
\cite{Nielsen2010}. These approaches provide filter guarantees in their
respective access models. Here the phase-oracle call is simply the uncontrolled
commuting evolution \(\exp(-\ii T H)\): for a diagonal two-body Ising
Hamiltonian it is exact, Trotter-free, and consists of \(O(n^2)\) commuting
rotations. The trade-off is that the selected band is generated by the spectrum
and the diffusion step, its peak estimates require statistical assumptions,
and Ising correlations can shift and reduce the resonance.

Phase-based versions of search have appeared in several contexts, including
continuous-time search, QAOA-inspired Grover circuits, arbitrary-phase and
fixed-point amplitude amplification, and physical implementations or
applications of Grover-type amplification
\cite{FarhiGutmann1998,Jiang2017,Hoyer2000,Yoder2014,Anikeeva2021,Sinitsyn2023,Nieman2024,Koch2022,Yan2022}.
These works show that search dynamics can be driven by phases rather than only
by ideal Boolean marks. The setting studied here is different in a specific
sense: the oracle is not a preconstructed marked-state oracle, nor is it built
by explicitly computing the Ising energy. Instead, the full Ising
Hamiltonian itself is used as a continuous spectral phase oracle. Every
computational-basis configuration receives a phase according to its energy, so
nearby energies are partially marked at the same time. In the arbitrary-phase,
fixed-point, and Gaussian amplitude-amplification constructions cited above, the
marked set or target transformation is still specified in advance. Here the
effective marked band instead emerges from the empirical spectrum and the
evolution time through the response derived below.

This continuous marking is the central issue. A Boolean oracle marks a sharply
defined subspace, whereas the Hamiltonian phase oracle marks an extended
spectral region with smoothly varying phases. It is therefore not obvious a
priori that repeated Grover diffusion reflections should convert this continuum
of phases into the same square-root amplification law as a Boolean marked-set
oracle. We show below that this does happen in a controlled regime: the
continuous phase oracle dynamically creates an effective resonant marked band,
with a well-defined center, width, height, and saturation time.

The algorithm alternates \(\ee^{-\ii T H}\) with the Grover diffusion reflection
about the uniform state. A spin configuration becomes resonant when
\begin{equation}
    T E_s\simeq \pi \pmod{2\pi}.
\end{equation}
The tunable time \(T\) therefore selects a target energy, while the diffusion
step converts the phase contrast near \(-1\) into amplitude growth. The
resonance has a finite but exponentially narrow width, so the task is naturally
an energy-selective target-energy search rather than broad energy-window
sampling. The useful operating regime is an intermediate high-density spectral
tail. In the central part of the spectrum the density of states is so large
that selecting a typical energy range is not a rare search problem; near the
extreme spectral edge the density is so sparse that smooth spectral information
cannot reliably place an exponentially narrow resonance. The high-density tail
lies between these limits: the target fraction is still exponentially small,
making direct sampling costly, but the local density remains large enough to
support a controlled resonant band. In this regime the Hamiltonian phase oracle
gives the same quadratic improvement as Grover search, in the sense that a
classical sampling cost proportional to the inverse target fraction is replaced
by a number of coherent oracle calls proportional to its square root. Here and
throughout, the classical baseline is independent uniform sampling of bit
strings followed by classical evaluation of their Ising energies. We do not
claim the same comparison with the best instance-specific rare-event, MCMC,
parallel-tempering, or optimization algorithms, whose performance depends on
landscape structure and whose thermal windows are generally much broader than
the exponentially narrow coherent resonance studied here.

The Ising--Grover idea was previously explored phenomenologically and
numerically as a method for finding low-energy Ising configurations
~\cite{PreviousPaper}. The present work develops the corresponding spectral
theory and shifts the emphasis from direct ground-state selection to controlled
amplification of a low-lying target-energy band.

\begingroup\sloppy
For the diagonal Ising Hamiltonian, the coefficients entering the
recurrence can be written as
\(g_m=2^{-n}\sum_s e^{-\ii mT E_s}
=Z(\ii mT)/Z(0)
=\langle\xi|e^{-\ii mTH}|\xi\rangle\),
where \(Z(\beta)=\operatorname{Tr}e^{-\beta H}\) and
\(|\xi\rangle=2^{-n/2}\sum_s|s\rangle\) is the uniform state.
Thus, \(g_m\) is the characteristic function of the uniformly weighted
energy spectrum evaluated at \(mT\). Through the generating function
\(G(z)=\sum_{m\geq0}g_m z^m\), these samples determine the spectral
response of the exact Grover-type recurrence. Closely related
characteristic functions occur in quantum work statistics, while the
connection between Loschmidt amplitudes and complex-time partition
functions is standard in dynamical-partition-function studies
\cite{Talkner2007,Heyl2013}.\par\endgroup

The first result of the paper is a spectral theory of this resonance. For an
annealed Gaussian density of states, appropriate for the average energy
distribution of a random Ising Hamiltonian with Gaussian couplings, we derive
the peak position, width, height, and saturation time. If the resonance contains
\(M\) configurations, the required number of Hamiltonian phase-oracle calls
scales as \(\Theta(\sqrt{2^n/M})\), reproducing the Grover-type square-root law
for the target-energy task. The second result concerns what is missed by a
smooth density-of-states or random-energy approximation. In a true Ising
landscape, energies of different configurations are correlated through their
spin overlaps. In particular, the averaging over the disorder gives $\langle E_i E_j \rangle \neq \langle E_i \rangle \langle E_j \rangle$.
These correlations enter the coherent characteristic function
\(g_m=2^{-n}\operatorname{Tr}\ee^{-\ii m T H}\), shift the resonance center by a
realization-dependent phase with \(\E\varphi_0=0\) and
\(\sqrt{\E(\varphi_0^2)}=O(n^{-3})\), and can also modify the
peak height. This shift is small on the scale of the full spectrum but large
compared with the exponentially narrow resonance width. It is therefore both an
algorithmic detuning and a phase-sensitive diagnostic of correlations beyond
the random-energy model.

The third result is that this detuning can be controlled for
prescribed-energy targeting. We show that spectral symmetrization removes the
odd phase response by construction, while an iterative calibration of \(T\) uses
measured output energies to retune the resonance of the original two-body
Hamiltonian. These procedures turn the correlation-induced shift from an
uncontrolled displacement into either a removable artifact or an experimentally
useful calibration signal.

The structural simplicity of the commuting oracle should not be confused
with near-term feasibility. The asymptotic regime requires an exponentially
long coherent iteration sequence and correspondingly small effective logical
errors. We therefore regard the algorithm as a fault-tolerant asymptotic
primitive, rather than a NISQ or generic analog proposal.

The paper is organized as follows. Section~\ref{sec:algorithm} defines the
Ising Hamiltonian phase oracle, the diffusion iteration, and the
energy-selective target problem. Section~\ref{sec:generating} derives the exact
recurrence and generating-function solution. Section~\ref{sec:gaussian}
analyzes the annealed Gaussian theory. Section~\ref{sec:ising} treats the
correlated random Ising spectrum and discusses the resonance shift as a probe of
spectral correlations. Section~\ref{sec:correction} presents targeting
strategies and resource estimates. Section~\ref{sec:numerics} gives numerical
validation, and Sec.~\ref{sec:conclusion} summarizes the physical implications.
Technical derivations are collected in the appendices.
\section{Algorithm and energy-selective target problem}
\label{sec:algorithm}

\subsection{Ising Hamiltonian as a phase oracle}

We consider \(n\) classical Ising spins \(s_i=\pm1\), encoded in computational-basis states \(|s\rangle=|s_1,\ldots,s_n\rangle\). The diagonal Ising Hamiltonian is
\begin{equation}
    H(s)=\sum_{i=1}^{n}h_i s_i +2\sum_{1\le i<j\le n}J_{ij}s_i s_j,
    \label{eq:ising-H}
\end{equation}
or, equivalently,
\begin{equation}
    H=\sum_{i=1}^{n}h_i Z_i +2\sum_{1\le i<j\le n}J_{ij}Z_iZ_j .
    \label{eq:ising-H-operator}
\end{equation}
The factor of \(2\) in the quadratic term is a normalization convention used throughout the analysis. The energy of configuration \(s\) is denoted by \(E_s=H(s)\). 

The couplings \(J_{ij}\) and fields \(h_i\) have units of energy, while
\(T\) has units of inverse energy. Throughout the paper we use the standard
deviation of the two-body couplings, \(\sigma_J\), as the energy unit.
Equivalently, writing \(J_{ij}=\sigma_J \widetilde J_{ij}\),
\(h_i=\sigma_J \widetilde h_i\), and \(\tau=\sigma_J T\), the oracle phase is
\[
    \ee^{-\ii T H}=\ee^{-\ii\tau \widetilde H},
    \qquad \widetilde H=H/\sigma_J .
\]
Thus only dimensionless products \(TJ_{ij}\) and \(T h_i\) enter the dynamics.
Unless stated otherwise, energies are quoted in units of \(\sigma_J\).

The initial state is the uniform superposition
\begin{equation}
    |\xi\rangle=\D^{-1/2}\sum_{s\in\{\pm1\}^n}|s\rangle,
    \qquad \D=2^n .
    \label{eq:xi}
\end{equation}
One iteration of the algorithm is
\begin{equation}
    W_T=D_\xi U_T,
    \qquad
    U_T=e^{-\ii T H},
    \qquad
    D_\xi=2|\xi\rangle\langle\xi|-I .
    \label{eq:iteration-operator}
\end{equation}
After \(K\) iterations,
\begin{equation}
    |\psi_K\rangle=W_T^K|\xi\rangle .
    \label{eq:psi-k}
\end{equation}
The operation \(U_T\) is the Hamiltonian phase oracle. Since all terms in Eq.~\eqref{eq:ising-H-operator} commute, it can be implemented exactly, without Trotter error, as
\begin{equation}
    U_T=
    \prod_{i=1}^{n}\ee^{-\ii T h_i Z_i}
    \prod_{i<j}\ee^{-2\ii T J_{ij} Z_iZ_j} .
    \label{eq:commuting-implementation}
\end{equation}
Thus a dense Ising instance requires \(O(n^2)\) one- and two-qubit phase rotations per application of \(U_T\), before accounting for the diffusion operation and fault-tolerant synthesis. This is an important physical feature of the proposal: the oracle is not a compiled comparison circuit but the native phase evolution of the problem Hamiltonian.

\subsection{Physical picture of the iteration}

It is useful to compare Eq.~\eqref{eq:iteration-operator} with the standard Grover iterate. In ordinary Grover search, a Boolean oracle applies a phase \(-1\) to the marked subspace and \(+1\) to the orthogonal complement. The diffusion operator then performs a reflection about the uniform state, leading to a rotation in the two-dimensional space spanned by marked and unmarked components.

Here the oracle is intrinsically continuous. It places every configuration on the unit circle according to its energy,
\begin{equation}
    |s\rangle\mapsto \ee^{-\ii T E_s}|s\rangle .
\end{equation}
There is therefore no sharply marked subspace before the dynamics starts. Configurations with phases very close to \(-1\) behave most like Grover-marked states, configurations farther away are only partially marked, and the rest of the spectrum contributes to the global phase background. In other words, the Hamiltonian oracle marks a whole energy neighborhood continuously rather than applying the same phase to all states in a Boolean target set.

This point is not merely semantic. With a continuum of phases, it is not obvious that repeated diffusion reflections should preserve coherent amplification over many iterations. Different detunings inside the energy neighborhood could in principle dephase, and the usual two-dimensional Grover rotation need not exist. The spectral recurrence derived below shows what replaces that picture: the diffusion step couples the energy-dependent amplitudes only through their spectral average, and repeated iterations generate a resonance kernel centered near
\begin{equation}
    T E_s\simeq \pi \pmod{2\pi}.
    \label{eq:phase-condition-intro}
\end{equation}
The kernel has a finite phase width, which decreases with the number of iterations. States inside this width are amplified coherently, while states outside it remain part of the nonresonant background. Thus the continuous oracle dynamically creates an effective marked set, rather than requiring one to be specified in advance.

For a prescribed positive target energy \(E_*\), the naive choice is \(T=\pi/E_*\). Because the phase is periodic, one must also be aware of possible resonances at other energies satisfying \(T E=(2m+1)\pi\). In the high-density tail regime considered below, the analysis focuses on the principal resonance at \(TE\simeq\pi\), and additional resonances can be avoided or filtered by the choice of target range.

This use of \(U_T\) is closely related to the Ising--Grover phenomenological construction of Ref.~\cite{PreviousPaper}, where the evolution time was chosen so that low- or high-energy states acquire an approximately Grover-like phase flip. In the present notation the same physical operation is treated as a tunable spectral phase oracle. The object selected by the algorithm is the finite resonance set defined below, and its behavior is characterized by a center, a width, and a peak height.

The finite width of the resonance is therefore not a defect but the physical resolution scale of the continuous oracle. A finite number of iterations cannot isolate an exact energy delta function. Instead, it selects all configurations whose phases lie within a narrow resonance region. For the asymptotic regime of interest this region is exponentially narrow in phase. Thus the algorithm is better described as energy-selective search rather than broad energy-window sampling.

\subsection{Target-energy resonance and success probability}

The selected set is defined by a small phase tolerance \(\delta\) and by a possible resonance shift \(\varphi_0\):
\begin{equation}
    \Rres(T,\delta,\varphi_0)
    =\left\{s:\left|T E_s-\pi-\varphi_0\right|\le \delta\right\}.
    \label{eq:resonance-set}
\end{equation}
In the annealed Gaussian theory \(\varphi_0=0\). For the correlated random
Ising ensemble we will find \(\E\varphi_0=0\) and
\(\sqrt{\E(\varphi_0^2)}=O(n^{-3})\). The success probability after \(K\) iterations is
\begin{equation}
    P_{\mathrm{succ}}(K,T;\Rres)=
    \sum_{s\in\Rres}\left|\langle s|\psi_K\rangle\right|^2 .
    \label{eq:success-prob}
\end{equation}
If the selected resonance contains \(M\) configurations and the evolution behaves like amplitude amplification on this effective set, then one expects \(O(\sqrt{\D/M})\) oracle calls. Here \(M\) is not supplied externally, as in a Boolean marked-set oracle. It is determined self-consistently by the local density of states and by the phase resolution of the iterate \(W_T\). The analysis below first computes this spectral resolution, then translates it into the effective number of selected configurations, and finally identifies the corrections caused by Ising spectral correlations.

We assume that the couplings \(J_{ij}\) are independent Gaussian random variables
with mean zero and variance \(\sigma_J^2\), and that the fields \(h_i\) are
independent Gaussian random variables with mean zero and variance \(\sigma_h^2\).
The choice \(\sigma_J=1\) used in the numerical work below is therefore only a
choice of energy units, not a nonzero mean coupling. For a fixed spin
configuration, the random energy is Gaussian. The annealed density of states is
therefore \cite{Parisi}
\begin{equation}
    f_G(E)=\frac{1}{\sqrt{2\pi}\Sigma_n}\exp\left(-\frac{E^2}{2\Sigma_n^2}\right),
    \label{eq:gaussian-density}
\end{equation}
with variance
\begin{equation}
    \Sigma_n^2=2n(n-1)\sigma_J^2+n\sigma_h^2 .
    \label{eq:Sigma-n}
\end{equation}
This annealed density correctly describes the average number of states in the energy interval, but it does not by itself describe correlations between different energy levels. Those correlations are treated in Sec.~\ref{sec:ising}.

The target-energy regime is specified by choosing \(E_*\) in a high-density tail. We parametrize the local mean level spacing by
\begin{equation}
    \Delta_n\sim n^{\mu}2^{-n/2},
    \label{eq:Delta-mu}
\end{equation}
through the condition
\begin{equation}
    2^n f_G(E_*)\Delta_n\sim1 .
    \label{eq:spacing-condition}
\end{equation}
Equivalently,
\begin{equation}
    \frac{E_*^2}{2\Sigma_n^2}
    =\frac{n}{2}\ln2+(\mu-1)\ln n+O(1).
    \label{eq:tail-energy}
\end{equation}
Thus \(E_*=O(n^{3/2})\) for fixed \(\sigma_J=O(1)\). This is far enough into the tail to make the search nontrivial, but not so far that the spectrum is reduced to isolated extreme levels separated by spacings of order unity.

In words, the algorithm targets an intermediate high-density tail of the Ising energy distribution. This region is very different from the central bulk, where the number of states per unit energy is exponentially large and a phase resonance would select a broad, highly populated part of the spectrum rather than a rare target set. It is also different from the extreme edge, where the density is so low that a smooth density-of-states estimate cannot reliably place an exponentially narrow resonance on an actual level. The high-density tail keeps both requirements in balance: the target states are rare enough that finding them by direct random sampling is exponentially hard, but the local density is still high enough that the continuous phase resonance contains a controlled, nonempty set of configurations. In this sense the effective marked set is not externally supplied; it is generated by the local spectral density and by the phase resolution of the iteration. The quadratic advantage is the usual amplitude-amplification advantage for this generated set: direct sampling from the uniform state needs a number of trials proportional to the inverse target fraction, whereas the Hamiltonian phase-oracle iteration needs the square root of that number of oracle calls.

\section{Exact spectral dynamics}
\label{sec:generating}

\subsection{Spectral recurrence}

The iteration depends on the Hamiltonian only through its spectrum in the computational basis. The empirical spectral measure is
\begin{equation}
    f(E)=\D^{-1}\sum_{s}\delta(E-E_s),
    \label{eq:empirical-density}
\end{equation}
and the state amplitudes can be grouped by energy as
\begin{equation}
    |\psi_K\rangle=\D^{-1/2}\sum_s a_K(E_s)|s\rangle,
    \qquad a_0(E)=1 .
    \label{eq:ak-definition}
\end{equation}
The amplitudes obey the exact recurrence
\begin{equation}
    a_K(E)=2\int \ee^{-\ii T\widetilde E}a_{K-1}(\widetilde E)f(\widetilde E)\,d\widetilde E
    -\ee^{-\ii T E}a_{K-1}(E) .
    \label{eq:ak-recurrence}
\end{equation}
The first term is the global average phase amplitude that is returned by the diffusion reflection, and the second term is the locally phase-rotated amplitude. Equation~\eqref{eq:ak-recurrence} is exact for a discrete spectrum and also defines a useful continuum approximation when \(f\) is replaced by a smooth density of states. The norm
\begin{equation}
    \int |a_K(E)|^2 f(E)\,dE=1
\end{equation}
is conserved because \(W_T\) is unitary.

The physical meaning of Eq.~\eqref{eq:ak-recurrence} is that the algorithm is a collective interference process in energy space. The energy-dependent phase \(\ee^{-\ii T E}\) attempts to separate resonant configurations from the rest of the spectrum, while the diffusion reflection couples every energy back to the spectral average. Thus a Grover-like phase-flip picture is recovered only after a local spectral reduction has been justified; it is not assumed at the outset. This is why the full density of states, and not only the target level, affects the amplification peak.

\subsection{Generating-function solution and spectral response}

The characteristic-function samples collect the energy phases that enter the diffusion average:
\begin{equation}
    g_m=\int \ee^{-\ii m T E}f(E)\,dE,
    \qquad m=0,1,2,\ldots,
    \label{eq:g_m_def}
\end{equation}
with \(g_0=1\). For the uniform initial state, \(g_m\) is simultaneously the characteristic
function of the empirical energy distribution, the normalized
imaginary-temperature partition function, and the uniform-state Loschmidt
amplitude. The generating function below turns these samples into the response
of the exact recurrence. The generating function is
\begin{equation}
    G(z)=\sum_{m=0}^{\infty}g_m z^m,
    \qquad |z|<1 .
    \label{eq:G-def}
\end{equation}
The scalar spectral average after the preceding iteration is
\begin{equation}
    s_K=\int \ee^{-\ii T E}a_{K-1}(E)f(E)\,dE,
    \qquad K\ge1,
    \label{eq:s-k-def}
\end{equation}
with \(s_0=1\). It satisfies
\begin{equation}
    s_K=2\sum_{m=0}^{K}(-1)^m g_m s_{K-m}-(-1)^K g_K,
    \label{eq:s-recurrence}
\end{equation}
and
\begin{equation}
    a_K(E) = (-1)^K\ee^{-\ii KET}
    \left[2\sum_{m=0}^{K}(-1)^m s_m\ee^{\ii mET} - 1\right].
    \label{eq:a-from-s}
\end{equation}
The ordinary generating function
\begin{equation}
    S(z)=\sum_{K=0}^{\infty}s_K z^K
\end{equation}
turns Eq.~\eqref{eq:s-recurrence} into
\begin{equation}
    S(-z)=\frac{G(z)}{2G(z)-1}
    =\frac12\left[1+\frac{1}{2G(z)-1}\right].
    \label{eq:S-G-relation}
\end{equation}
Substitution into the Cauchy representation of the coefficients yields, after \(K-1\) completed iterations,
\begin{widetext}
\begin{equation}
    a_{K-1}(E)=(-1)^{K-1}\ee^{-\ii(K-1)TE}\frac{1}{2\pi}
    \int_{-\pi}^{\pi}
    \frac{1-r^{-K}\ee^{\ii K(TE-\pi-\varphi)}}
         {1-r^{-1}\ee^{\ii(TE-\pi-\varphi)}}
    \frac{d\varphi}{2G(-r\ee^{\ii\varphi})-1} .
    \label{eq:a-solution-main}
\end{equation}
\end{widetext}
Here \(0<r<1\) is an auxiliary contour radius; it is not an algorithmic parameter. For estimates at iteration number \(K\), one chooses \(1-r=O(1/K)\), which keeps \(r^{-K}=O(1)\).

A derivation of Eq.~\eqref{eq:a-solution-main} is given in Appendix~\ref{app:generating}. Equation~\eqref{eq:a-solution-main} is the central analytic expression of the paper. The first factor under the integral is a finite-time resonance kernel peaked at
\begin{equation}
    \varphi=TE-\pi
\end{equation}
with width \(O(1/K)\). Keeping the generating function in this empirical form turns the Hamiltonian-oracle idea into a spectral-response theory: replacing \(G\) by its smooth Gaussian approximation gives the annealed resonance of Sec.~\ref{sec:gaussian}, whereas keeping the correlated Ising characteristic function gives the displacement and peak-height corrections of Sec.~\ref{sec:ising}. The second factor is the inverse of the spectral response
\begin{equation}
    C_r(\varphi)=2G(-r\ee^{\ii\varphi})-1 .
    \label{eq:C-def}
\end{equation}
The denominator \(C_r\) plays the role of a susceptibility of the spectral ensemble to the Grover reflection. A small value of \(C_r\) means that a phase component near \(\varphi\) is resonantly amplified by repeated reflections. Near the relevant resonance, the local response may be written as
\begin{equation}
    C_r(\varphi)\simeq \eps_{\eff}+\ii \kappa_C(\varphi-\varphi_0),
    \label{eq:local-C}
\end{equation}
and the amplification peak is centered at
\begin{equation}
    TE-\pi-\varphi_0=0,
    \label{eq:peak-position}
\end{equation}
The peak has phase width of order \(\eps_{\eff}/|\kappa_C|\) and reaches amplitude height of order \(1/\eps_{\eff}\) in the smooth local-response approximation. This local response form is what connects the microscopic spectrum to the observable search performance.

\section{Annealed Gaussian spectral theory}
\label{sec:gaussian}

The annealed Gaussian density \eqref{eq:gaussian-density} is the simplest model for the Ising density of states. Its characteristic-function samples are
\begin{equation}
    g_m=\exp\left[-\frac{(m\Sigma_n T)^2}{2}\right].
    \label{eq:gaussian-gm}
\end{equation}
The dimensionless Gaussian width parameter is
\begin{equation}
    a=\frac{(\Sigma_n T)^2}{2}.
    \label{eq:a-gaussian}
\end{equation}
In the high-density tail regime, \(T=\pi/E_*\) and \(a\ll1\). As shown in Appendix~\ref{app:gaussian}, the local Gaussian response is
\begin{equation}
    C_1^{(G)}(\varphi)
    \simeq \eps_G+\ii \kappa_C\varphi,
    \qquad
    \eps_G=2\sqrt{\frac{\pi}{a}}\exp\left(-\frac{\pi^2}{4a}\right).
    \label{eq:gaussian-local-C}
\end{equation}
Substituting this form into Eq.~\eqref{eq:a-solution-main} gives the resonant peak shape
\begin{equation}
    a_{K-1}(x)\simeq
    \frac{1}{|\kappa_C|}
    \frac{1-\exp[K(-\gamma+\ii x)]}
         {1-\exp[-\gamma+\ii x]},
    \qquad x=TE-\pi,
    \label{eq:gaussian-peak-shape}
\end{equation}
up to an overall phase and nonresonant corrections. Here
\begin{equation}
    \gamma=\frac{\eps_G}{|\kappa_C|}\simeq 2\eps_G .
\end{equation}
On resonance, the amplitude grows linearly with \(K\) for \(1\ll K\ll1/\gamma\) and saturates at \(a_{\mathrm{sat}}\simeq1/(|\kappa_C|\gamma)=1/\eps_G\) for \(K\gg1/\gamma\), to leading order in the narrow-linewidth limit. The phase width is \(O(\gamma)\), and the corresponding energy resolution is
\begin{equation}
    \delta E_{\peak}\sim \frac{\gamma}{T}.
    \label{eq:energy-width}
\end{equation}
The distinction between phase width and energy width is essential: the phase resonance is narrow, but its conversion into energy depends on the target time \(T\).

We tune the principal resonance to a target energy \(E_*\) by setting
\begin{equation}
    T_*=\frac{\pi}{E_*}.
    \label{eq:T-star}
\end{equation}
Using Eqs.~\eqref{eq:tail-energy} and \eqref{eq:gaussian-local-C}, the Gaussian phase width \(\gamma\) is proportional to \(\eps_G\), which scales as
\begin{equation}
    \eps_G\sim n^{3/2-\mu}2^{-n/2} .
    \label{eq:epsilon-scaling}
\end{equation}
Consequently,
\begin{equation}
    K_{\peak}^{(G)}
    \sim \frac{1}{\eps_G}
    \sim n^{\mu-3/2}2^{n/2} .
    \label{eq:Amax-G}
\end{equation}
The number of energy levels inside the selected resonance is obtained by dividing the energy resolution by the mean level spacing. Using \(\gamma=\Theta(\eps_G)\),
\begin{equation}
    M_{\peak}
    \sim \frac{\delta E_{\peak}}{\Delta_n}
    \sim \frac{\gamma}{T_*\Delta_n}
    \sim n^{3-2\mu} .
    \label{eq:Mpeak}
\end{equation}
Thus \(\mu=3/2\) selects \(O(1)\) configurations on average, while \(\mu<3/2\) selects a polynomially large set. For \(\mu>3/2\), a randomly placed resonance typically contains no level unless additional spectral information is used to tune \(T\).

Combining Eqs.~\eqref{eq:Amax-G} and \eqref{eq:Mpeak},
\begin{equation}
    K_{\peak}^{(G)}
    \sim n^{\mu-3/2}2^{n/2}
    \sim \sqrt{\frac{2^n}{M_{\peak}}} .
    \label{eq:Gaussian-Grover-scaling}
\end{equation}
This is the Grover-type scaling law of the annealed theory. The derivation also fixes what the effective marked set means in the Hamiltonian-oracle setting: the selected cardinality \(M_{\peak}\) is generated by the phase resonance itself. In the special case \(M_{\peak}=O(1)\), the result reduces to the familiar \(K=O(2^{n/2})\) scaling, but the target set is still defined by the spectral response rather than by an externally specified Boolean predicate. Direct random sampling would require \(O(2^n/M_{\peak})\) energy evaluations to hit the same target-energy resonance, whereas the Hamiltonian phase-oracle algorithm requires the square root of this number of iterations.

At this stopping scale the resonant amplitude is
\(A_{\peak}=\Theta(K_{\peak}^{(G)})\). Therefore, in the annealed Gaussian
regime,
\begin{equation*}
    P_{\mathrm{succ}}(K_{\peak}^{(G)})
    \sim \frac{M_{\peak}}{2^n}\left(K_{\peak}^{(G)}\right)^2
    =\Theta(1).
\end{equation*}
Thus Eq.~\eqref{eq:Gaussian-Grover-scaling} does not hide an additional
exponential number of repetitions. This constant-success statement is specific
to the annealed regime; the correlated peak-height correction discussed below
can reduce the target weight.

The local exponent \(\mu\), and hence the amplification scale, need not be
known in advance. The plateau described by
Eq.~\eqref{eq:gaussian-peak-shape} applies to the smooth annealed response,
appropriate to a resonant band containing many levels. In this regime a
constant-factor overshoot does not erase the resonance. For a discrete sparse
band with \(M_{\peak}=O(1)\), as at \(\mu=3/2\), a near-resonant spin-flip
pair can instead produce Grover-type oscillations, and a fixed iteration count
can fall near a node of the success probability.

To accommodate this sparse regime, we use a randomized BBHT-style schedule
\cite{Boyer1998}: at stage \(j\), draw \(K\) uniformly from
\(\{0,\ldots,L_j-1\}\), where \(L_j=\lceil b^jL_0\rceil\) grows
geometrically with \(b>1\), and verify the measured energy classically.
Averaging over \(K\) avoids a deterministic schedule repeatedly selecting
oscillation nodes. Provided this averaged success probability is bounded below
by a constant beyond the amplification scale, a suitable constant number of
independent runs per stage gives expected total iteration count
\(O(K_{\peak})\). The plateau itself is not assumed in the sparse regime.

The physical content of this result is the following. In the spectral bulk there are too many nearly resonant states, so the algorithm produces broad energy-selective sampling rather than isolation of a small set. At the extreme edge there are too few levels, so the resonance is unlikely to contain the desired state without detailed spectral information. Between these regimes lies the high-density tail, where the density of states is low enough for energy selectivity but high enough for the phase resonance to contain a controlled number of configurations. This is the natural operating regime of the present primitive.

\section{Correlated random Ising spectra}
\label{sec:ising}

\subsection{Why the annealed density is not enough}

For a fixed spin configuration, the random Ising energy is Gaussian with variance \eqref{eq:Sigma-n}. However, the algorithm is controlled not only by the one-point density of energies but by the characteristic function of the entire empirical spectrum. This is where the physics of the spin-glass energy landscape enters. A mean density of states, or a mean estimate of an edge energy, fixes the coarse location of a target phase but not the coherent detuning of a particular disorder realization. Two configurations with a large spin overlap have strongly correlated energies because they share many products \(s_i s_j\). Configurations with small overlap are less correlated. A random-energy model ignores this structure and therefore misses effects that are visible in the generating function.

To expose these correlations we focus on the zero-field ensemble, \(h_i=0\).
The linear fields can be included as an additional Gaussian factor and do not
change the leading effects discussed here when \(\sigma_h=O(\sigma_J)\). In the dimensionless units introduced above,
\(\widetilde J_{ij}=J_{ij}/\sigma_J\sim\mathcal N(0,1)\); in the numerical
simulations we set \(\sigma_J=1\), so \(J_{ij}\) itself has unit variance. The
characteristic-function coefficients are
\begin{equation}
    g_m=2^{-n}\sum_{s\in\{\pm1\}^n}
    \exp\left[-2\ii mT\sum_{i<j}J_{ij}s_is_j\right] .
    \label{eq:gm-ising}
\end{equation}
Averaging over the couplings gives
\begin{equation}
    \E g_m=\exp[-\alpha n(n-1)m^2],
    \label{eq:E-gm}
\end{equation}
which coincides with the Gaussian annealed prediction. Near \(T\approx T_*\) we have
\begin{equation}
    \alpha=(\sigma_J T)^2 = \frac{\pi^2}{2n^3\ln2}
    \left[1-\frac{2(\mu-1)}{\ln2}\frac{\ln n}{n}+O\left(\frac1n\right)\right].
    \label{eq:alpha-def}
\end{equation}
In the numerical simulations reported below we set \(\sigma_J=1\), so that
\(\alpha=T^2\) and all energies are dimensionless in units of \(\sigma_J\).

The second moment is different. Direct averaging, with details given in Appendix~\ref{app:covariance}, gives
\begin{widetext}
\begin{equation}
    \E\left[g_m g_\ell^*\right]
    =2^{-n}\exp\left\{-\alpha\left[n(n-1)(m^2+\ell^2)+2nm\ell\right]\right\}
    \sum_{q=0}^{n}\binom{n}{q}\exp\left[2\alpha m\ell(n-2q)^2\right].
    \label{eq:gm-gl-covariance}
\end{equation}
\end{widetext}
The exponent has been written in expanded form to make the overall factor
\(n\) explicit. In particular, setting \(\ell=0\) immediately recovers
Eq.~\eqref{eq:E-gm}.
The integer \(q\) counts the Hamming distance between two spin configurations, or equivalently their overlap. This formula is a compact expression of the correlated Ising landscape. In the regime of large \(m\ell\), the terms \(q=0\) and \(q=n\), corresponding to identical or globally flipped spin configurations, give
\begin{equation}
    \E\left[g_m g_\ell^*\right]
    \simeq 2^{-(n-1)}\exp[-\alpha n(n-1)(m-\ell)^2] .
    \label{eq:large-ml-covariance}
\end{equation}
This term represents a persistent noise floor of order \(2^{-n/2}\) in the sequence \(g_m\). It is small, but it accumulates in the generating function near the unit circle and influences the resonance.

\subsection{Correlation-induced displacement of the target resonance}

The resonance position is determined by the imaginary part of
\begin{equation}
    C_r(\varphi)=2G(-r\ee^{\ii\varphi})-1 .
\end{equation}
For the Gaussian response, \(\operatorname{Im}C_r(\varphi)\simeq \kappa_C\varphi\) with \(\kappa_C\simeq-1/2\). In a typical Ising realization, spectral correlations add a small random offset. The estimates derived from Eq.~\eqref{eq:gm-gl-covariance} give
\begin{equation}
    \E\left\{\left(\operatorname{Im}\left[G(-r\ee^{\ii\varphi})-G_0(-r\ee^{\ii\varphi})\right]\right)^2\right\}
    \sim n^{-6}+\varphi^2 n^{-4},
    \label{eq:imag-difference-main}
\end{equation}
and
\begin{equation}
    \E\left\{\left(\frac{d}{d\varphi}\operatorname{Im}\left[G(-r\ee^{\ii\varphi})-G_0(-r\ee^{\ii\varphi})\right]\right)^2\right\}
    \sim n^{-4}+\varphi^2 n^{-6}.
    \label{eq:imag-derivative-main}
\end{equation}
Here \(G_0\) is the annealed Gaussian generating function. These bounds imply the local form
\begin{equation}
    \operatorname{Im}C_r(\varphi)=\kappa_C(\varphi-\varphi_0)+O((\varphi-\varphi_0)^2),
    \qquad
    \varphi_0=O(n^{-3})
    \label{eq:phi0-main}
\end{equation}
within the local linearization. More explicitly, they imply \(\E(\varphi_0^2)=O(n^{-6})\), so the displacement is \(O(n^{-3})\) in root-mean-square.

The selected energy is therefore not determined by the naive equation \(TE=\pi\) but by
\begin{equation}
    T E-\pi-\varphi_0=0 .
    \label{eq:shifted-peak-condition}
\end{equation}
The energy displacement is
\begin{equation}
    \delta E_{\mathrm{shift}}
    =\frac{E_*\varphi_0}{\pi}
    =O(n^{-3/2}) .
    \label{eq:energy-shift}
\end{equation}
This is a physically important scale separation. The displacement is tiny compared with the total spectral width, which is \(O(n)\), and even compared with the target energy, which is \(O(n^{3/2})\). But it is large compared with the exponentially narrow resonance width $\delta E_\mathrm{peak} = O(n^{3-\mu}2^{-n/2})$ selected by the algorithm. Therefore a phase shift that would be negligible in ordinary spectral-density estimates becomes decisive for energy-selective quantum search.

\subsection{Resonance shift as a probe of spectral correlations}
\label{subsec:diagnostic_shift}

The displacement \(\varphi_0\) is not only a nuisance for targeting. It is also a coherent diagnostic of the energy landscape. The annealed density of states describes how many configurations occur in the energy interval, but the phase-oracle dynamics depends on the complex sums
\begin{equation}
    g_m=2^{-n}\operatorname{Tr}\ee^{-\ii m T H}
    =2^{-n}Z(\beta=\ii mT),
    \label{eq:gm-imaginary-temperature}
\end{equation}
which are normalized partition functions at imaginary inverse temperature. Such quantities are sensitive to interference between phases from different spin configurations. Two spectra with nearly the same density of states can therefore have different coherent responses if their level correlations differ.

In this sense the target resonance functions as a spectroscopic probe. In the annealed Gaussian theory the imaginary part of \(C_r(\varphi)\) crosses zero at \(\varphi=0\), and the naive condition \(TE=\pi\) is correct. For a correlated Ising spectrum the zero is shifted to \(\varphi_0\). A batch of uncalibrated runs can provide an estimator \(\widehat E_{\mathrm{out}}\) of the amplified peak center, with uncertainty determined by the sampled output distribution and the batch size. It then gives
\begin{equation}
    \widehat\varphi_0\approx T\widehat E_{\mathrm{out}}-\pi.
    \label{eq:phi0-from-output}
\end{equation}
Thus the same detuning that complicates prescribed-energy search can be used to estimate the correlation-induced response of a particular Ising instance.

This observation separates three levels of description. The annealed Gaussian density fixes the coarse scale of the target energy and the expected level spacing. The random-energy approximation would treat the individual energies as essentially independent samples from that density. The Ising Hamiltonian, however, has an overlap structure encoded in Eq.~\eqref{eq:gm-gl-covariance}. This equation identifies the ensemble-statistical origin of the shift: the covariance of the empirical characteristic function is indexed by Hamming distance and hence by spin overlap. For a fixed realization, however, the uniform-start dynamics depends only on the unordered empirical spectrum through the coefficients \(g_m\). We therefore interpret the shift as a probe of sample-specific spectral fluctuations whose ensemble statistics originate in the overlap structure, rather than as a reconstruction of the assignment of individual energies to Hamming neighborhoods. The shift is small in ordinary spectral units, but coherent amplification resolves phases on an exponentially fine scale; consequently a polynomially small shift becomes observable and algorithmically relevant. Future implementations of Hamiltonian phase oracles could use this effect as a calibration signal or as a diagnostic for how far a given Ising instance is from a random-energy description.

\subsection{Peak-height correction from the real response}

The real part of \(C_r\) controls the height of the amplification peak. In the Gaussian theory it is \(\eps_G\). In the correlated Ising ensemble it receives realization-dependent corrections.
The detailed derivation is given in Appendix~\ref{app:asymptotics}. Under the explicit typical-value assumption stated there, the second-moment calculation gives the following heuristic asymptotic estimate:
\begin{equation}
    A_{\max}^{(\mathrm{Ising})}
    \sim
    \min\left[2^{n/2},\;
    n^{\mu-3/2}2^{n/2},\;
    2^{(1-d_*)n/2}
    n^{-\frac12+\frac{\mu-1}{1+(1-2s_*)^2}}
    \right],
    \label{eq:Amax-Ising}
\end{equation}
up to powers of \(n\), where $s_*\approx 0.308$, $d_* \approx 0.01936$. This is a typical-value estimate based on a second moment, not a rigorous probabilistic bound. The contour radius \(r=1-2^{-n/2}\) used in obtaining it is an auxiliary steepest-descent choice within the exact contour representation \eqref{eq:a-solution-main}; it is not a physical control or an algorithmic tuning parameter.

The first term is the natural maximum amplification for a discrete Hilbert space of size \(2^n\). The second term is the ideal Gaussian prediction. The third term is a correlation-induced Ising correction. It is very difficult to observe at the system sizes accessible to exact numerical simulations, but it is important conceptually: a statement of ideal \(2^{n/2}\) amplification is justified in the annealed Gaussian theory, whereas the correlated Ising ensemble requires a more careful qualification. The correlated landscape does not merely shift the resonance; it can also broaden or damp the spectral response.

\section{Targeting a prescribed energy}
\label{sec:correction}

For prescribed-energy targeting, the correlated spectrum creates a
resonance-center displacement in addition to the possible peak-height
reduction. The naive choice \(T=\pi/E_*\) can therefore amplify energies near,
but not centered on, the intended target. We discuss two ways to correct this
displacement.

\subsection{Spectral symmetrization}

The random displacement \(\varphi_0\) is caused by the imaginary part of the response denominator. A direct way to remove it is to symmetrize the spectrum. Introduce an ancillary qubit and define
\begin{equation}
    \widetilde H=Z_a\otimes H .
    \label{eq:sym-H}
\end{equation}
The extended search starts from
\(|\widetilde\xi\rangle=|+\rangle_a\otimes|\xi\rangle\), where
\(|+\rangle_a=(|0\rangle_a+|1\rangle_a)/\sqrt{2}\) is the \(+1\)
eigenstate of \(X_a\). It uses the reflection
\(D_{\widetilde\xi}=2|\widetilde\xi\rangle
\langle\widetilde\xi|-I\), so the two ancilla sectors enter the
spectral measure with equal weights.
If \(H|E_j\rangle=E_j|E_j\rangle\), then
\begin{equation}
    \widetilde H |\pm\rangle_z\otimes |E_j\rangle
    =\pm E_j |\pm\rangle_z\otimes |E_j\rangle .
\end{equation}
Here \(|+\rangle_z=|0\rangle_a\) and
\(|-\rangle_z=|1\rangle_a\) are the two eigenstates of \(Z_a\).
The spectral distribution becomes
\begin{equation}
    \widetilde f(E)=\frac12[f(E)+f(-E)] .
    \label{eq:sym-density}
\end{equation}
Consequently,
\begin{equation}
    \widetilde g_m
    =\int \ee^{-\ii m T E}\widetilde f(E)\,dE
    =\int \cos(mTE)f(E)\,dE,
    \label{eq:sym-gm}
\end{equation}
which is real. The imaginary part of \(\widetilde C_r(\varphi)\) is then odd and vanishes at \(\varphi=0\), so the resonance is centered at the naive condition \(TE=\pi\).

The physical price is that the Hamiltonian is no longer purely two-body:
\begin{equation}
    \widetilde H=
    \sum_i h_i Z_aZ_i
    +2\sum_{i<j}J_{ij}Z_aZ_iZ_j .
    \label{eq:sym-H-expanded}
\end{equation}
In a digital implementation the three-body phase gates can be decomposed into elementary two-qubit gates using standard ancilla-mediated constructions. A standard parity decomposition of each \(\exp(-\ii\theta Z_aZ_iZ_j)\) uses four CNOT gates and one single-qubit \(R_Z\) rotation, two CNOTs more than the corresponding two-body \(ZZ\) phase gadget. This is an \(O(1)\) overhead per Ising term and preserves the \(O(n^2)\) oracle scaling. The remedy is not generically native in analog hardware, where a direct or engineered three-body coupling would be required. The symmetrized algorithm also amplifies both signs of the target energy, since the spectrum contains \(\pm E_j\). This is not a fundamental obstacle: the sign can be checked after measurement by classical evaluation of the original Ising energy.

\subsection{Iterative calibration of the evolution time}

A second strategy keeps the original Ising Hamiltonian and treats \(T\) as a calibratable physical control parameter. For a desired energy \(E_*\), the correct time satisfies
\begin{equation}
    T=\frac{\pi+\varphi_0(T)}{E_*} .
    \label{eq:T-fixed-point}
\end{equation}
This gives a spectral interpretation to the time fine tuning used in low-energy applications of the same Ising--Grover iterate \cite{PreviousPaper}: the quantity being tuned is the zero of \(\operatorname{Im}C_r\), not only the mean phase-flip condition.  
Starting from \(T_1=\pi/E_*\), at update \(k\) one performs \(N_{\mathrm{cal}}\) independent preparations, evaluates the Ising energy of every measured bit string, and obtains an estimator \(\widehat E_k\) for the center of the amplified output peak. This gives
\begin{equation}
    \widehat{\varphi}_0(T_k)\simeq T_k\widehat E_k-\pi .
    \label{eq:phi-estimate}
\end{equation}
The next trial value is
\begin{equation}
    T_{k+1}=\frac{\pi+\widehat{\varphi}_0(T_k)}{E_*},
    \label{eq:T2-update}
\end{equation}
and the procedure is repeated. The batch estimate makes explicit that a single measured configuration is not, by itself, a reliable estimate of the peak center.
For a sparse resonant band, the iteration count is randomized within each
calibration batch as in Sec.~\ref{sec:gaussian}; using only a fixed \(K\)
could suppress the peak signal by sampling an oscillation node.

This calibration loop is experimentally natural: it uses only measured bit strings and the known classical Ising energy function, and it does not require reconstructing the whole spectrum. The contraction estimate follows from implicit differentiation of
\begin{equation}
    \operatorname{Im}C_r(\varphi_0(T),T)=0 .
    \label{eq:implicit-phi0}
\end{equation}
Near the resonance, \(\partial_\varphi\operatorname{Im}C_r\simeq -1/2\), while the random correction to the \(T\) derivative has root-mean-square scale \(O(n^{-3/2})\). Hence
\begin{equation}
    \frac{d\varphi_0}{dT}=O(n^{-3/2}),
    \qquad
    \frac{d}{dT}\left[\frac{\pi+\varphi_0(T)}{E_*}\right]
    =O(n^{-3}) .
    \label{eq:contraction}
\end{equation}
The map is therefore contracting with a polynomially small contraction factor. Reaching an exponentially small phase tolerance \(2^{-n/2}\) requires
\begin{equation}
    k_{\mathrm{tune}}=O\left(\frac{n}{\ln n}\right)
    \label{eq:tuning-iterations}
\end{equation}
updates, up to constants and measurement noise. Equation~\eqref{eq:tuning-iterations}
therefore gives only the polynomial number of feedback updates. The total
calibration cost also contains the statistical number \(N_{\mathrm{cal}}\) of
preparations per update, which is retained explicitly below rather than assumed
to be polynomial.

\subsection{\texorpdfstring{Precision and coherence requirements}{Precision and coherence requirements}}

The narrow resonance imposes separate requirements on repeatable detuning and
on errors that vary between iterations. After calibration, let \(T_0\) place
\(E_*\) at the peak and write \(T=T_0+\delta T\). From
Eq.~\eqref{eq:gaussian-peak-shape}, the finite-time response to the fixed phase
detuning \(x=E_*\delta T\) is
\begin{equation*}
    a_{K-1}(x)\simeq
    \frac{1}{|\kappa_C|}
    \frac{1-\exp[K(-\gamma+\ii x)]}
         {1-\exp[-\gamma+\ii x]},
    \qquad \gamma=\frac{\eps_{\eff}}{|\kappa_C|},
\end{equation*}
within the smooth local-response approximation.
The target remains within the final linewidth provided
\begin{equation*}
    |E_*\delta T|\lesssim\gamma,
    \qquad
    \frac{|\delta T|}{T_0}\lesssim\frac{\gamma}{\pi}
    \sim n^{3/2-\mu}2^{-n/2}.
\end{equation*}
This finite-time width already includes the repeated action of a fixed
detuning over the full sequence; it should therefore not be divided by
\(K_{\peak}\) a second time. Since \(K_{\peak}=O(\gamma^{-1})\), a repeatable
common phase offset has tolerance \(O(K_{\peak}^{-1})\) and can in principle be
removed by the calibration loop above.

Iteration-to-iteration phase noise has a different accumulation law. For
independent zero-mean Gaussian phase kicks \(\delta\phi_j\) with per-step
variance \(v_\phi\), coherence between contributions separated by \(d\) steps
is attenuated by \(\exp(-d v_\phi/2)\). Equivalently, the random-phase model
adds a dephasing scale of order \(v_\phi\) to the linewidth. Retaining an
order-one fraction of the resonant signal therefore requires
\begin{equation*}
    v_\phi=O(K_{\peak}^{-1}),
    \qquad
    \phi_{\mathrm{rms}}=O(K_{\peak}^{-1/2}),
\end{equation*}
consistent with the known noisy-Grover scaling \cite{Shenvi2003}. This noise is
not removed by calibration, and tolerances to nonuniform or temporally
correlated noise are model dependent.

The static requirement corresponds to \(\Theta(n)\) bits of logical timing or
angle precision. In a fault-tolerant digital implementation, standard
\(Z\)-rotation synthesis has cost logarithmic in the inverse target error, so
this precision contributes polynomial synthesis overhead
\cite{RossSelinger2016}. The number of coherent iterations itself remains
exponential, making the protocol an asymptotic fault-tolerant primitive rather
than a NISQ proposal.

\subsection{Resource scaling and practical interpretation}

For a dense two-body Ising Hamiltonian, one Hamiltonian phase-oracle call requires \(O(n^2)\) commuting \(ZZ\) rotations. The diffusion operation is implementable by Hadamards around a multi-controlled phase and has polynomial cost in \(n\). Therefore, in the Gaussian spectral regime, the gate-level cost of producing a state in the selected resonance is
\begin{equation}
    O\left(n^2\sqrt{\frac{2^n}{M_{\peak}}}\right)
    \label{eq:gate-cost-gaussian}
\end{equation}
This is a per-preparation count after the target time has been calibrated.
This estimate counts the leading number of commuting \(ZZ\) phase rotations. Connectivity routing, fault-tolerant synthesis, logical error correction, and the precision requirements discussed above contribute polynomial overheads and are not included in the leading exponential comparison. 
A calibration with \(k_{\mathrm{tune}}\) updates and \(N_{\mathrm{cal}}\) preparations per update has the separate one-time cost
\begin{equation*}
    O\!\left(N_{\mathrm{cal}}k_{\mathrm{tune}}K_{\peak}\right)
    \quad\text{oracle calls},
    \qquad
    O\!\left(N_{\mathrm{cal}}k_{\mathrm{tune}}n^2K_{\peak}\right)
    \quad\text{Ising phase rotations}.
\end{equation*}
The required \(N_{\mathrm{cal}}\) depends on the desired confidence and on the realized output distribution; it is not included in Eq.~\eqref{eq:gate-cost-gaussian}.

This scaling should be interpreted as an oracle-level and circuit-level statement for target-energy search, not as a proof that arbitrary Ising optimization becomes easy. The algorithm still has exponential complexity in \(n\) when \(M_{\peak}=O(1)\). Its significance is the square-root reduction relative to the stated independent-uniform-sampling baseline under the stated spectral assumptions, together with the physically natural implementation of the phase oracle. If the correlation-induced estimate in Eq.~\eqref{eq:Amax-Ising} limits the peak height, then additional repetitions or more refined phase matching are required. This limitation originates in the correlated Ising spectrum, not in the cost of simulating the Hamiltonian evolution.

\section{Numerical validation}
\label{sec:numerics}

The numerical calculations test the physical picture developed above: the iteration should produce a narrow resonance at the predicted phase, the annealed Gaussian model should describe the coarse peak shape, and a discrete random Ising realization should display a realization-dependent displacement of the resonance. The energy-resolved amplitudes plotted below determine \(P_{\mathrm{succ}}\) through Eq.~\eqref{eq:success-prob} by summing \(\D^{-1}|a_K(E_s)|^2\) over the selected resonance set. Thus the peak position, width, and height shown in the figures are spectral diagnostics of the success probability.

All random-Ising simulations use dimensionless units with \(\sigma_J=1\).
Unless stated otherwise, the couplings are sampled as
\(J_{ij}\sim\mathcal N(0,1)\), the fields are set to zero, \(h_i=0\), and the
factor of \(2\) in Eq.~\eqref{eq:ising-H} is kept. Thus the simulation time
\(T\) is the dimensionless time \(\tau=\sigma_J T\).

\begin{figure}[ht]
\centering
\includegraphics[width=0.9\linewidth]{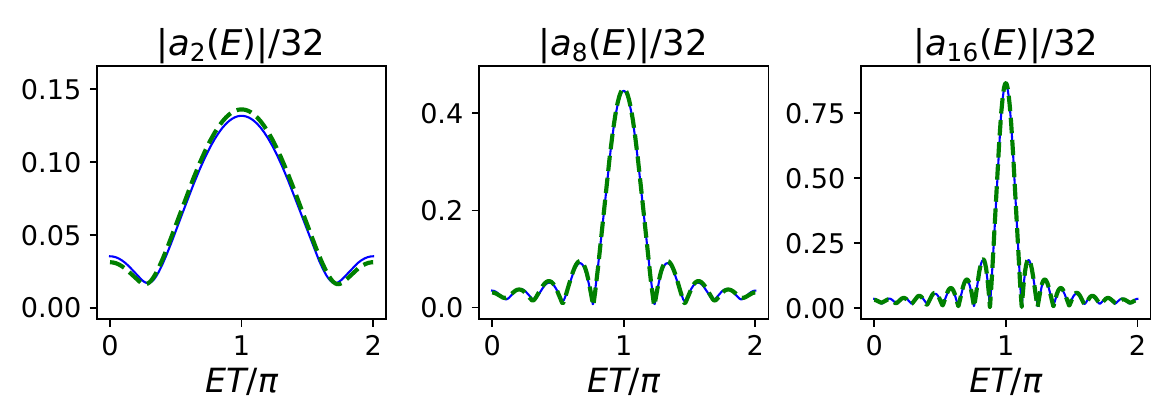}
\caption{
Amplification profile for an annealed Gaussian spectrum near the principal
phase resonance \(ET=\pi\). The horizontal axis is the dimensionless phase
\(ET/\pi\), and the three panels show the amplitude magnitude after
\(K=2,8,16\) iterations. Solid curves are obtained from the exact spectral
recurrence, while dashed curves show Eq.~\eqref{eq:gaussian-peak-shape},
with \(K+1\) substituted for \(K\) to represent \(K\) completed iterations.
Both are plotted as \(|a_K|/32\), using a common normalization factor.
The Gaussian width is \(\Sigma T=0.72\). The agreement near
\(ET/\pi=1\) illustrates that the local response captures the growth and
narrowing of the Gaussian resonance.
}
\label{fig:gaussian_solution}
\end{figure}

Figure~\ref{fig:gaussian_solution} illustrates the Gaussian resonance in the
phase variable \(ET/\pi\). For \(\Sigma T=0.72\), the local response
approximation reproduces the peak region well, while the oscillatory tails are
finite-time interference fringes of the resonance kernel.

\begin{figure}[ht]
\centering
\includegraphics[width=0.85\linewidth]{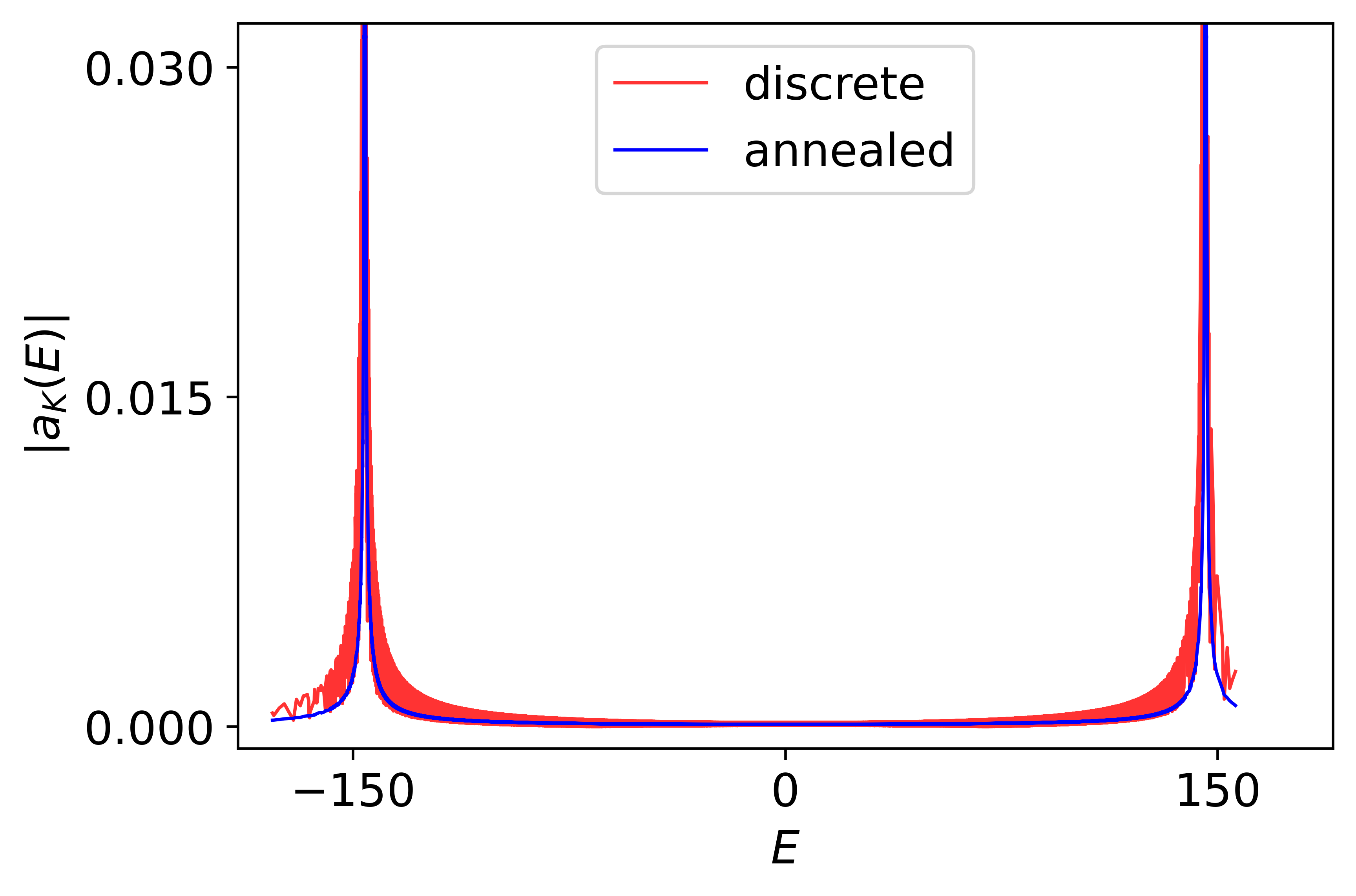}
\caption{
Energy-space amplification profile for a single zero-field random Ising
instance compared with the annealed Gaussian prediction. The Ising instance has
\(n=24\), \(h_i=0\), and independent couplings \(J_{ij}\sim\mathcal N(0,1)\).
The red curve shows the response of the discrete Ising spectrum, while the blue
curve shows the annealed Gaussian model with the same energy variance. Both
profiles develop resonances near \(ET=\pm\pi\), but the discrete Ising response
is shifted and distorted by spectral correlations.
}
\label{fig:ising_solution_1}
\end{figure}

Figure~\ref{fig:ising_solution_1} compares a discrete random Ising spectrum
with the annealed Gaussian density for \(n=24\). Both responses develop
amplification near \(ET=\pm\pi\), showing that the physical resonance mechanism
survives the replacement of the smooth density by a correlated discrete
spectrum. The comparison also illustrates the limitation of a purely annealed
description: the Ising peak is shifted relative to the Gaussian peak.

\begin{figure}[ht]
\centering
\includegraphics[width=0.95\linewidth]{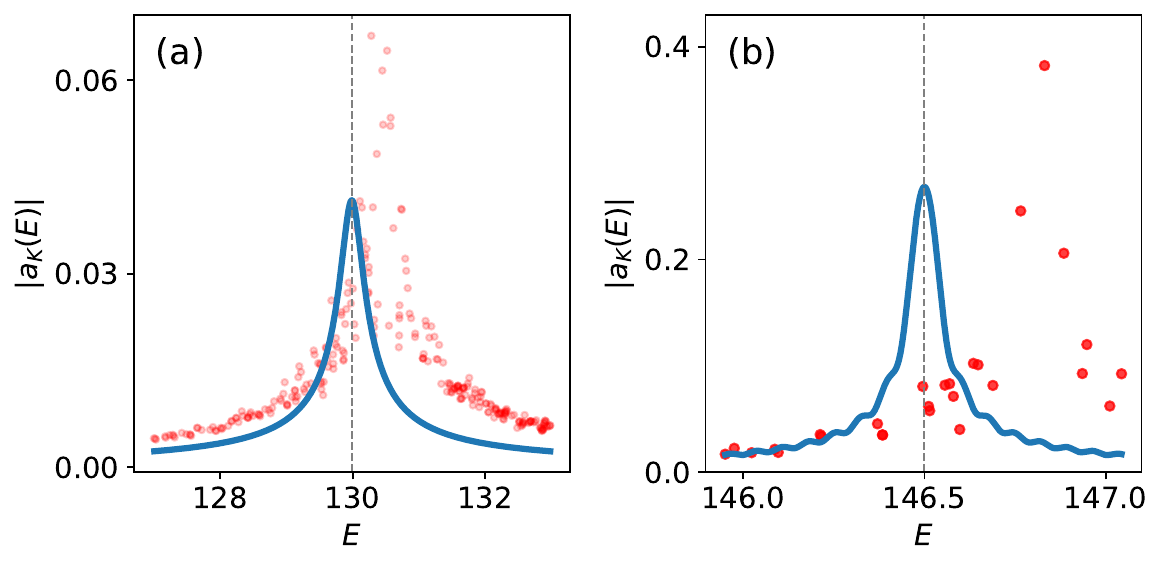}
\caption{
Zoom of the positive-energy resonance for zero-field random Ising spectra at
\(n=24\). Blue curves show the annealed Gaussian response, red points show the
amplitudes on the discrete Ising energy levels, and the vertical dashed line
marks the Gaussian target energy \(E_*=\pi/T\). Panel (a) corresponds to a
denser spectral tail, approximately \(\mu\simeq0.6\), with a broader and lower
peak. Panel (b) corresponds to a sparser tail, approximately \(\mu\simeq1.2\),
with a narrower and higher peak. In both cases the discrete Ising peak is
displaced relative to the Gaussian target, illustrating the realization-dependent
phase shift \(\varphi_0\).
}
\label{fig:ising_solution_2}
\end{figure}

Figure~\ref{fig:ising_solution_2} zooms in on the resonance for two target energies. In the denser tail, corresponding approximately to \(\mu\approx0.6\), the peak is broader and lower. In the sparser tail, approximately \(\mu\approx1.2\), it is sharper and higher, as predicted by the scaling \(A_{\max}\sim n^{\mu-3/2}2^{n/2}\). The relative size of the displacement compared with the asymptotic Gaussian peak width scales as approximately \(n^{\mu-9/2}2^{n/2}\), up to finite-size corrections and prefactors. At the accessible system size \(n=24\), the observed shift should therefore be interpreted as a finite-size manifestation of the scale separation rather than as a quantitative test of the asymptotic power law.

The \(d_*\)-dependent asymptotic reduction of the peak height in
Eq.~\eqref{eq:Amax-Ising} is too small to be reliably observed at these system
sizes. The simulations therefore support the leading physical conclusions: energy-selective amplification occurs, the Gaussian theory captures the basic resonant peak, and correlations in the Ising spectrum produce a measurable displacement of the target resonance.

\subsection{Scaling of the correlation-induced displacement}
\begin{figure}[t]
\centering
\includegraphics[width=0.95\linewidth]{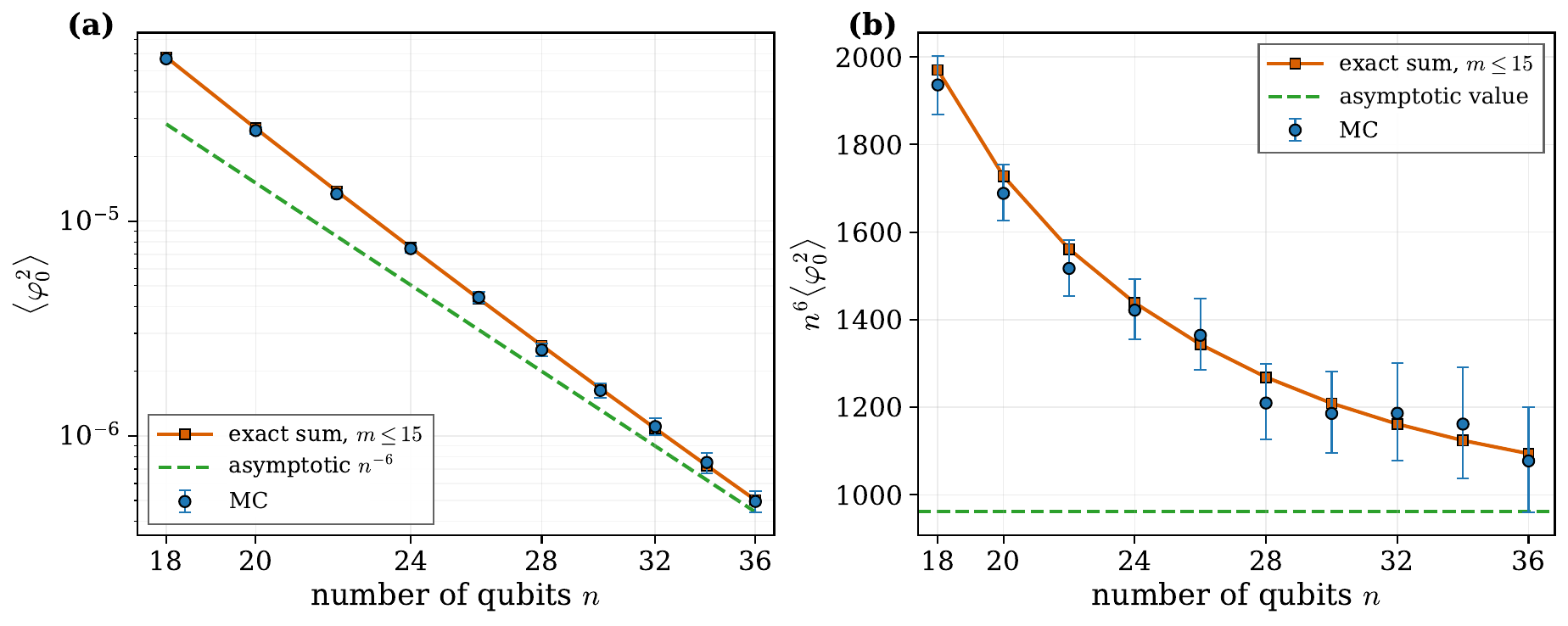}
\caption{
Scaling of the correlation-induced resonance displacement in the zero-field
random Ising ensemble. The simulations use \(\sigma_J=1\),
\(J_{ij}\sim\mathcal N(0,1)\), \(h_i=0\), \(\mu=3/2\), and the truncated
characteristic function with \(m\le15\). Panel (a) shows the disorder-averaged
second moment \(\langle\varphi_0^2\rangle\) for \(n=18,\ldots,36\). Circles are Monte Carlo estimates, with error bars denoting 95\% bootstrap
confidence intervals over disorder realizations. Monte Carlo sampling is over
disorder realizations, not over spin configurations. Squares show the finite-\(n\) covariance sum evaluated with the same truncation.
The dashed line in panel (a) is the leading asymptotic prediction
\(C_\infty n^{-6}\), with \(C_\infty=\pi^6/[3(\ln2)^3]\).
Panel (b) shows \(n^6\langle\varphi_0^2\rangle\), and its horizontal dashed
line is \(C_\infty\simeq962.28\).
}
\label{fig:phi0_scaling}
\end{figure}

To test the scaling prediction for the correlation-induced displacement,
we computed the truncated characteristic-function coefficients
\(g_m=2^{-n}\Tr\ee^{-\ii m T H}\), \(m\le 15\), for random Ising instances
with \(n=18,\ldots,36\). For each disorder realization the trace over the
\(2^n\) spin configurations was evaluated deterministically, not sampled: we
split the spins into two equal halves, performed the resulting block
contractions on a GPU, and used the zero-field spin-flip symmetry
\(H(s)=H(-s)\) to reduce the sum by a factor of two. Thus the Monte Carlo
average in Fig.~\ref{fig:phi0_scaling} is only over independent disorder
realizations. For each instance we extracted the local displacement from the
zero of the imaginary part of the response function
\(C(\varphi)=2\sum_{m=0}^{15}g_m(-1)^m\ee^{\ii m\varphi}-1\),
using the linear estimate
\[
\varphi_0\simeq
-\frac{\operatorname{Im}C(0)}
{\partial_\varphi\operatorname{Im}C(0)} .
\]
The disorder average of \(\varphi_0\) is consistent with zero.
The second moment agrees with the finite-\(n\) covariance prediction. Figure~\ref{fig:phi0_scaling}
compares the Monte Carlo estimate of \(\langle\varphi_0^2\rangle\) with the
finite-\(n\) covariance sum obtained from Eq.~\eqref{eq:gm-gl-covariance}.
The leading smooth-response asymptote is
\[
\langle\varphi_0^2\rangle
=\frac{C_\infty}{n^6}
 \left[1+O\!\left(\frac{\ln n}{n}\right)\right],
\qquad C_\infty=\frac{\pi^6}{3(\ln2)^3}\simeq962.28,
\]
as shown in Appendix~\ref{app:asymptotics}.
The dashed lines show \(C_\infty n^{-6}\) in panel (a) and \(C_\infty\)
in panel (b). The finite-size data still show a drift of
\(n^6\langle\varphi_0^2\rangle\); they have not reached a plateau.

\clearpage
\section{Discussion and conclusion}
\label{sec:conclusion}

The principal result of this work is an exact spectral-response theory for the
iterate \(D_\xi\exp(-\ii TH)\). The energy-grouped recurrence and the contour
representation separate a finite-time resolution kernel from the response
\(C_r(\varphi)=2G(-r\exp(\ii\varphi))-1\), built from the characteristic-function
samples \(g_m\). This response fixes the resonance center and height for an
empirical diagonal spectrum. The Grover-type scaling then follows in the
annealed high-density-tail regime rather than being assumed through a predefined
Boolean marked set.

The exact generating-function representation \eqref{eq:a-solution-main} makes the mechanism transparent. The finite-time kernel determines the phase resolution set by the number of iterations, while the spectral response \(2G(-r\ee^{\ii\varphi})-1\) determines where amplification occurs and how high the peak can grow. In the annealed Gaussian model this response is almost singular at the phase \(-1\), leading to a resonance of phase width \(\gamma\simeq2\eps_G\) and height \(1/\eps_G\). For target energies in a high-density tail with local mean spacing \(\Delta_n\sim n^\mu2^{-n/2}\), the selected resonance contains \(M_{\peak}\sim n^{3-2\mu}\) configurations and is reached after \(O(\sqrt{2^n/M_{\peak}})\) Hamiltonian phase-oracle calls. Its success probability is \(\Theta(1)\) at saturation in the annealed theory. This is a square-root query improvement over independent uniform sampling of the same narrow band, not a claim of generic advantage over all structure-exploiting classical algorithms.

The correlated Ising spectrum adds two physically important effects. First, the resonance center is shifted by a realization-dependent phase with \(\E\varphi_0=0\) and \(\sqrt{\E(\varphi_0^2)}=O(n^{-3})\). Although this shift is small in ordinary spectral units, it is large compared with the exponentially narrow resonance width and must be corrected for prescribed-energy search. Second, correlations can modify the real part of the spectral response and reduce the asymptotic peak height relative to the ideal Gaussian prediction. This is the sense in which the random Ising landscape is not equivalent to an independent random-energy model for the present algorithm.

The same effect gives the algorithm a diagnostic interpretation. Since \(g_m=2^{-n}\operatorname{Tr}\ee^{-\ii mTH}\) is a complex-temperature spectral response, the resonance shift probes sample-specific spectral fluctuations that are invisible in the annealed density of states and whose ensemble covariance is generated by Ising overlaps. An uncalibrated run can therefore be viewed as a coherent measurement of the detuning of a particular Ising instance from the random-energy prediction. It does not reconstruct which energies belong to particular Hamming neighborhoods. This turns a potential drawback of Hamiltonian phase-oracle search into a useful signal: the algorithm not only searches near a prescribed energy, but also exposes phase-sensitive structure of the Ising landscape.

The two correction strategies developed here turn the energy displacement into a controllable effect. Spectral symmetrization removes the odd part of the response by doubling the spectrum with an ancillary qubit. It is conceptually clean and fixes the resonance position, but it converts two-body Ising terms into three-body terms. Iterative calibration of \(T\) preserves the original Hamiltonian and uses measured energies as feedback. It is therefore closer to an experimental calibration protocol: the device is run, bit strings are measured, their classical energies are evaluated, and the phase resonance is retuned. The number of feedback updates is polynomial, while the statistical sampling cost is retained separately through \(N_{\mathrm{cal}}\). Thus the resonance-center shift can be removed by symmetrization or compensated by calibration; neither procedure removes the separate peak-height correction.

The algorithm should not be viewed as a universal solution of Ising optimization or as a generic exact ground-state finder. Its controlled regime is target-energy search in a spectral region where the density of states is high enough for the resonance to contain configurations but low enough for the search to be selective. This regime is physically relevant for sampling rare configurations in spin-glass tails, benchmarking Hamiltonian-based oracles, and exploring alternatives to threshold-based Boolean search circuits. The extreme spectral edge, hardware noise, and possible phase-matched variants that compensate both the imaginary and real parts of the correlated response remain important directions for future work.

A warm start \(|\eta\rangle=\sum_s c_s|s\rangle\) changes the spectral measure,
characteristic-function samples, and diffusion reflection to
\begin{equation*}
    f_\eta(E)=\sum_s|c_s|^2\delta(E-E_s),
    \qquad g_m^{(\eta)}=\langle\eta|\exp(-\ii mTH)|\eta\rangle,
    \qquad D_\eta=2|\eta\rangle\langle\eta|-I.
\end{equation*}
A larger initial target-band overlap can reduce the amplification count, but the
response becomes state dependent and the preparation and reflection costs must
be included. The present uniform-state asymptotics therefore do not transfer
unchanged; we leave this extension for future work.

More broadly, the analysis shows that Grover-type speedup is not restricted to
an ideal Boolean oracle with a sharply marked subspace. A Hamiltonian oracle
that marks computational-basis states continuously through energy-dependent
phases can dynamically form an effective resonant marked band and amplify it
with the same square-root scaling, provided the target band lies in the
controlled high-density spectral-tail regime. In this sense, energy-selective
Hamiltonian phase oracles provide a continuous spectral realization of
Grover-type amplification rather than merely a physical implementation of a
predefined Boolean oracle.

\section*{Data Availability}

The data and code supporting the numerical results are available in
Zenodo~\cite{Phi0ZenodoData}. The archive contains the scripts
used to generate the characteristic-function coefficients
\(g_m=2^{-n}\Tr\ee^{-\ii m T H}\) for the zero-field random
Ising ensemble, the NumPy data files used in Fig.~\ref{fig:phi0_scaling}, and
the scripts used to compute \(\varphi_0\), bootstrap confidence intervals,
finite-\(n\) covariance-sum comparisons, and the plotted figures. The random
Ising data are stored as non-pickled \texttt{.npz} archives and use the
dimensionless convention \(\sigma_J=1\), \(J_{ij}\sim\mathcal N(0,1)\), and
\(h_i=0\).

\appendix

\section{Derivation of the generating-function formula}
\label{app:generating}

Starting from Eq.~\eqref{eq:ak-recurrence}, define \(s_K\) by Eq.~\eqref{eq:s-k-def}. Then
\begin{equation}
    a_K(E)=2s_K-\ee^{-\ii T E}a_{K-1}(E).
\end{equation}
Iterating this relation gives
\begin{equation}
    a_K(E)=(-1)^K\ee^{-\ii KTE}
    \left[2\sum_{m=0}^{K}(-1)^m s_m\ee^{\ii mTE}-1\right].
    \label{eq:app-ak-sm}
\end{equation}
Multiplying by \(\ee^{-\ii TE}\), integrating against \(f(E)\), and using the definition of \(g_m\) gives Eq.~\eqref{eq:s-recurrence}. Introducing
\begin{equation}
    S(z)=\sum_{K=0}^{\infty}s_K z^K,
    \qquad
    G(z)=\sum_{m=0}^{\infty}g_m z^m,
\end{equation}
we find
\begin{equation}
    S(-z)=2G(z)S(-z)-G(z),
\end{equation}
which gives Eq.~\eqref{eq:S-G-relation}. The Cauchy formula for \((-1)^m s_m\) on a circle \(|\alpha|=r<1\) yields
\begin{equation}
    2\sum_{m=0}^{K-1}(-1)^m z^m s_m-1
    =\frac{1}{2\pi\ii}\oint
    \frac{1-(z/\alpha)^K}{1-z/\alpha}
    \frac{1}{2G(\alpha)-1}\frac{d\alpha}{\alpha} .
\end{equation}
Setting \(z=\ee^{\ii TE}\) and \(\alpha=-r\ee^{\ii\varphi}\) gives Eq.~\eqref{eq:a-solution-main}.

\section{Gaussian peak from Poisson summation}
\label{app:gaussian}

For the Gaussian density,
\begin{equation}
    g_m=\ee^{-a m^2},
    \qquad a=\frac{(\Sigma T)^2}{2}.
\end{equation}
The real part of \(C(0)=2G(-1)-1\) is
\begin{equation}
    C(0)=\sum_{m=-\infty}^{\infty}(-1)^m\ee^{-a m^2}.
\end{equation}
Poisson summation gives
\begin{equation}
    \sum_{m=-\infty}^{\infty}(-1)^m\ee^{-a m^2}
    =\sqrt{\frac{\pi}{a}}\sum_{q=-\infty}^{\infty}
    \exp\left[-\frac{\pi^2(q+1/2)^2}{a}\right].
\end{equation}
For \(a\ll1\), the terms \(q=0\) and \(q=-1\) dominate, giving Eq.~\eqref{eq:gaussian-local-C}. The slope of the imaginary part follows from
\begin{equation}
    \frac{d}{d\varphi}\operatorname{Im}G(-\ee^{\ii\varphi})\bigg|_{\varphi=0}
    =\sum_{m=0}^{\infty}(-1)^m m\ee^{-a m^2},
\end{equation}
whose Abel-regularized limit as \(a\to0\) is \(-1/4\). Therefore the slope of \(2G-1\) is \(-1/2\).

\section{Covariance of the Ising characteristic function}
\label{app:covariance}

For the purely quadratic Hamiltonian,
\begin{equation}
    g_m=2^{-n}\sum_s
    \exp\left[-2\ii mT\sum_{i<j}J_{ij}s_is_j\right].
\end{equation}
Averaging over \(J_{ij}\sim\mathcal N(0,\sigma_J^2)\) gives
\begin{equation}
    \E g_m=2^{-n}\sum_s
    \exp\left[-2m^2T^2\sigma_J^2\sum_{i<j}(s_is_j)^2\right]
    =\exp[-\alpha n(n-1)m^2].
\end{equation}
For the covariance,
\begin{align}
    \E[g_mg_\ell^*]
    &=2^{-2n}\sum_{s,t}
    \exp\left[-2\sigma_J^2T^2\sum_{i<j}(m s_is_j-\ell t_it_j)^2\right] .
\end{align}
Let \(u_i=s_it_i\). Then
\begin{equation}
    \sum_{i<j}s_is_jt_it_j=\sum_{i<j}u_iu_j
    =\frac12\left[\left(\sum_i u_i\right)^2-n\right].
\end{equation}
Counting configurations by the number \(q\) of \(u_i=-1\) gives Eq.~\eqref{eq:gm-gl-covariance}.

\section{Asymptotic estimates for the random generating function and peak height}
\label{app:asymptotics}

This appendix records the estimates used in Sec.~\ref{sec:ising}. From Eq.~\eqref{eq:gm-gl-covariance}, the mean square difference between the imaginary parts of the random and annealed generating functions can be written as
\begin{multline}
\E\left\{\left(\operatorname{Im}\left[G(-r\ee^{\ii\varphi})-G_0(-r\ee^{\ii\varphi})\right]\right)^2\right\}
=    
\\
\frac{1}{4}\sum_{m,\ell=-\infty}^{\infty} e^{-\alpha n(n-1)(m^2+\ell^2)}r^{|m|+|\ell|}(-1)^{m+\ell}\ee^{\ii\varphi(m-\ell)}
    \left[2^{-n}\sum_{q=0}^{n}\binom{n}{q}\ee^{2\alpha m\ell[(n-2q)^2-n]}-1
    \right]\sgn(m\ell).
\end{multline}
The corresponding expression for the real part is the same without the factor \(\sgn(m\ell)\). The terms \(q=0,n\) behave as \(2^{-n}\ee^{-\alpha n(n-1)(m-\ell)^2}\), and the factor \(r<1\) is required for convergence of the sums near the unit circle.

The leading coefficient of the smooth-response displacement variance follows
from the cubic term in the odd part of Eq.~\eqref{eq:gm-gl-covariance}.
For the binomially distributed overlap, set \(X=(n-2q)^2-n\). Then
\(\E X=0\) and \(\E X^3=48\binom n3\), giving
\[
\operatorname{Cov}(\operatorname{Im}g_m,\operatorname{Im}g_\ell)
\simeq64\alpha^3\binom n3 m^3\ell^3\ee^{-a(m^2+\ell^2)},
\qquad a=\alpha n(n-1).
\]
Using the full Gaussian-regulated sum
\(S_3(a)=\sum_{m\ge1}(-1)^m m^3\ee^{-am^2}=1/8+O(a)\)
and \(\kappa_C=-1/2+O(a)\), the local linearization yields
\[
\langle\varphi_0^2\rangle
\simeq\frac{256\alpha^3\binom n3}{\kappa_C^2}S_3(a)^2
=16\alpha^3\binom n3[1+O(n^{-1})].
\]
For \(\mu=3/2\), Eq.~\eqref{eq:alpha-def} gives
\(\alpha=\pi^2/[2(\ln2)n^3]\,[1+O(\ln n/n)]\), and hence
\[
\langle\varphi_0^2\rangle
=\frac{\pi^6}{3(\ln2)^3n^6}
 \left[1+O\!\left(\frac{\ln n}{n}\right)\right].
\]

The estimates \eqref{eq:imag-difference-main} and \eqref{eq:imag-derivative-main} follow from boundedness, for \(|\varphi|<\varphi_{\max}<\pi\) and \(0<a<1\), of the alternating sums
\begin{equation}
\sum_{m=0}^{\infty}(-1)^m m^{2k+1}\cos(\varphi m)\ee^{-a m^2}
\end{equation}
and of the corresponding sine sums divided by \(\varphi\). The absence of \(\sgn(m\ell)\) in the real part leads to the exponentially small contributions. For \(|\varphi|=o(1/n)\), one obtains
\begin{widetext}
\begin{align}
    \E\left\{\left(\operatorname{Re}\left[G(-r\ee^{\ii\varphi})-G_0(-r\ee^{\ii\varphi})\right]\right)^2\right\}
    \sim{}&
    \frac{1}{1-r}n^{3/2-\mu}2^{-3n/2}
    +(1-r)^2n^{-4}
    +(1-r)n^{3/2-\mu}2^{-n/2}
    \nonumber\\
    &+2^{-n(1-d_*)}n^{1+\frac{2(1-\mu)}{1+(1-2s_*)^2}} .
    \label{eq:real-difference-main}
\end{align}
\end{widetext}
The number \(s_*\simeq0.308\) in \eqref{eq:real-difference-main} maximizes the function
\begin{equation}
    F(s)=-s\log_2s-(1-s)\log_2(1-s)
    -\frac{2s(1-s)}{1+(1-2s)^2},
    \label{eq:F-s-def}
\end{equation}
with \(d_*\) defined by
\begin{equation}
    F(s_*)=\frac12+d_* ,
    \qquad d_*\simeq0.01936 .
    \label{eq:dstar-def}
\end{equation}
The first two terms arise from \(q=0,n\). The third term comes from the first correction in \(1-r\) for \(q\ne0,n\). The last term comes from \(q\ne0,n\) with \(r=1\). After double Poisson summation over \(m,\ell\) and parametrisation $q=sn$ we get integral $\int P(n,s)2^{nF(s)}ds$ and estimate it by saddle-point.

\begingroup\sloppy
For the following estimate we make the explicit typical-value assumption that the random variable $\operatorname{Re}\left[G(-r\ee^{\ii\varphi})-G_0(-r\ee^{\ii\varphi})\right]$ has a typical magnitude of the order of its root-mean-square value,\par\endgroup
\begin{equation}
\varepsilon(r)=\sqrt{\E\left\{\left(\operatorname{Re}\left[G(-r\ee^{\ii\varphi})-G_0(-r\ee^{\ii\varphi})\right]\right)^2\right\}},
\end{equation}
which was obtained in \eqref{eq:real-difference-main}. No concentration theorem is established here; consequently, the peak-height result based on this substitution is a heuristic typical-value estimate rather than a rigorous probabilistic bound.

The first factor in \eqref{eq:a-solution-main} limits the magnitude of the gain by the parameter K. From the condition $r^{-K} \sim 1$, we find that the amplification is limited by the value $1/(1-r)$.

An individual level near the resonance gives a contribution
\begin{equation}
\frac{1}{2^n}\sum_{k=0}^{\infty}(-r)^k
 \ee^{\ii k(\varphi-ET)}\sim\frac{1}{2^n(1-r)}
\end{equation}
to $G(-r\ee^{\ii\varphi})$. The average amplification  cannot be affected by a single energy level and we conclude that the amplification cannot be larger than $2^n(1-r)$.

Our analysis shows that the peak height is given by
\begin{equation}
    A_{\max}^{(\mathrm{Ising})}
    \sim
    \min\left[\frac{1}{\eps_{\eff}(r)},\frac{1}{1-r},2^n(1-r)\right] .
    \label{eq:Amax-general-ising}
\end{equation}
Optimizing the analytical contour estimate gives $r=1-2^{-n/2}$ and Eq.~\eqref{eq:Amax-Ising}. This choice of $r$ is part of the asymptotic evaluation of the exact contour integral and is not a parameter of the quantum algorithm.


\begin{thebibliography}{49}%
\makeatletter
\providecommand \@ifxundefined [1]{%
 \@ifx{#1\undefined}
}%
\providecommand \@ifnum [1]{%
 \ifnum #1\expandafter \@firstoftwo
 \else \expandafter \@secondoftwo
 \fi
}%
\providecommand \@ifx [1]{%
 \ifx #1\expandafter \@firstoftwo
 \else \expandafter \@secondoftwo
 \fi
}%
\providecommand \natexlab [1]{#1}%
\providecommand \enquote  [1]{``#1''}%
\providecommand \bibnamefont  [1]{#1}%
\providecommand \bibfnamefont [1]{#1}%
\providecommand \citenamefont [1]{#1}%
\providecommand \href@noop [0]{\@secondoftwo}%
\providecommand \href [0]{\begingroup \@sanitize@url \@href}%
\providecommand \@href[1]{\@@startlink{#1}\@@href}%
\providecommand \@@href[1]{\endgroup#1\@@endlink}%
\providecommand \@sanitize@url [0]{\catcode `\\12\catcode `\$12\catcode `\&12\catcode `\#12\catcode `\^12\catcode `\_12\catcode `\%12\relax}%
\providecommand \@@startlink[1]{}%
\providecommand \@@endlink[0]{}%
\providecommand \url  [0]{\begingroup\@sanitize@url \@url }%
\providecommand \@url [1]{\endgroup\@href {#1}{\urlprefix }}%
\providecommand \urlprefix  [0]{URL }%
\providecommand \Eprint [0]{\href }%
\providecommand \doibase [0]{https://doi.org/}%
\providecommand \selectlanguage [0]{\@gobble}%
\providecommand \bibinfo  [0]{\@secondoftwo}%
\providecommand \bibfield  [0]{\@secondoftwo}%
\providecommand \translation [1]{[#1]}%
\providecommand \BibitemOpen [0]{}%
\providecommand \bibitemStop [0]{}%
\providecommand \bibitemNoStop [0]{.\EOS\space}%
\providecommand \EOS [0]{\spacefactor3000\relax}%
\providecommand \BibitemShut  [1]{\csname bibitem#1\endcsname}%
\let\auto@bib@innerbib\@empty
%</preamble>
\bibitem [{\citenamefont {Sherrington}\ and\ \citenamefont {Kirkpatrick}(1975)}]{SK}%
  \BibitemOpen
  \bibfield  {author} {\bibinfo {author} {\bibfnamefont {D.}~\bibnamefont {Sherrington}}\ and\ \bibinfo {author} {\bibfnamefont {S.}~\bibnamefont {Kirkpatrick}},\ }\bibfield  {title} {\bibinfo {title} {Solvable model of a spin-glass},\ }\href {https://doi.org/10.1103/PhysRevLett.35.1792} {\bibfield  {journal} {\bibinfo  {journal} {Phys. Rev. Lett.}\ }\textbf {\bibinfo {volume} {35}},\ \bibinfo {pages} {1792} (\bibinfo {year} {1975})}\BibitemShut {NoStop}%
\bibitem [{\citenamefont {Derrida}(1981)}]{Derrida1981}%
  \BibitemOpen
  \bibfield  {author} {\bibinfo {author} {\bibfnamefont {B.}~\bibnamefont {Derrida}},\ }\bibfield  {title} {\bibinfo {title} {Random-energy model: An exactly solvable model of disordered systems},\ }\href {https://doi.org/10.1103/PhysRevB.24.2613} {\bibfield  {journal} {\bibinfo  {journal} {Phys. Rev. B}\ }\textbf {\bibinfo {volume} {24}},\ \bibinfo {pages} {2613} (\bibinfo {year} {1981})}\BibitemShut {NoStop}%
\bibitem [{\citenamefont {Fan}\ and\ \citenamefont {McCoy}(1969)}]{Fan1969}%
  \BibitemOpen
  \bibfield  {author} {\bibinfo {author} {\bibfnamefont {C.}~\bibnamefont {Fan}}\ and\ \bibinfo {author} {\bibfnamefont {B.~M.}\ \bibnamefont {McCoy}},\ }\bibfield  {title} {\bibinfo {title} {One-dimensional {Ising} model with random exchange energy},\ }\href {https://doi.org/10.1103/PhysRev.182.614} {\bibfield  {journal} {\bibinfo  {journal} {Phys. Rev.}\ }\textbf {\bibinfo {volume} {182}},\ \bibinfo {pages} {614} (\bibinfo {year} {1969})}\BibitemShut {NoStop}%
\bibitem [{\citenamefont {Lucas}(2014)}]{Lucas2014}%
  \BibitemOpen
  \bibfield  {author} {\bibinfo {author} {\bibfnamefont {A.}~\bibnamefont {Lucas}},\ }\bibfield  {title} {\bibinfo {title} {{Ising} formulations of many {NP} problems},\ }\href {https://doi.org/10.3389/fphy.2014.00005} {\bibfield  {journal} {\bibinfo  {journal} {Front. Phys.}\ }\textbf {\bibinfo {volume} {2}},\ \bibinfo {pages} {5} (\bibinfo {year} {2014})}\BibitemShut {NoStop}%
\bibitem [{\citenamefont {Kadowaki}\ and\ \citenamefont {Nishimori}(1998)}]{KadowakiNishimori1998}%
  \BibitemOpen
  \bibfield  {author} {\bibinfo {author} {\bibfnamefont {T.}~\bibnamefont {Kadowaki}}\ and\ \bibinfo {author} {\bibfnamefont {H.}~\bibnamefont {Nishimori}},\ }\bibfield  {title} {\bibinfo {title} {Quantum annealing in the transverse {Ising} model},\ }\href {https://doi.org/10.1103/PhysRevE.58.5355} {\bibfield  {journal} {\bibinfo  {journal} {Phys. Rev. E}\ }\textbf {\bibinfo {volume} {58}},\ \bibinfo {pages} {5355} (\bibinfo {year} {1998})}\BibitemShut {NoStop}%
\bibitem [{\citenamefont {Farhi}\ \emph {et~al.}(2001)\citenamefont {Farhi}, \citenamefont {Goldstone}, \citenamefont {Gutmann}, \citenamefont {Lapan}, \citenamefont {Lundgren},\ and\ \citenamefont {Preda}}]{AnnealingFarhi2001}%
  \BibitemOpen
  \bibfield  {author} {\bibinfo {author} {\bibfnamefont {E.}~\bibnamefont {Farhi}}, \bibinfo {author} {\bibfnamefont {J.}~\bibnamefont {Goldstone}}, \bibinfo {author} {\bibfnamefont {S.}~\bibnamefont {Gutmann}}, \bibinfo {author} {\bibfnamefont {J.}~\bibnamefont {Lapan}}, \bibinfo {author} {\bibfnamefont {A.}~\bibnamefont {Lundgren}},\ and\ \bibinfo {author} {\bibfnamefont {D.}~\bibnamefont {Preda}},\ }\bibfield  {title} {\bibinfo {title} {A quantum adiabatic evolution algorithm applied to random instances of an {NP}-complete problem},\ }\href {https://doi.org/10.1126/science.1057726} {\bibfield  {journal} {\bibinfo  {journal} {Science}\ }\textbf {\bibinfo {volume} {292}},\ \bibinfo {pages} {472} (\bibinfo {year} {2001})}\BibitemShut {NoStop}%
\bibitem [{\citenamefont {Albash}\ and\ \citenamefont {Lidar}(2018)}]{AlbashLidar2018}%
  \BibitemOpen
  \bibfield  {author} {\bibinfo {author} {\bibfnamefont {T.}~\bibnamefont {Albash}}\ and\ \bibinfo {author} {\bibfnamefont {D.~A.}\ \bibnamefont {Lidar}},\ }\bibfield  {title} {\bibinfo {title} {Adiabatic quantum computation},\ }\href {https://doi.org/10.1103/RevModPhys.90.015002} {\bibfield  {journal} {\bibinfo  {journal} {Rev. Mod. Phys.}\ }\textbf {\bibinfo {volume} {90}},\ \bibinfo {pages} {015002} (\bibinfo {year} {2018})}\BibitemShut {NoStop}%
\bibitem [{\citenamefont {Mohseni}\ \emph {et~al.}(2022)\citenamefont {Mohseni}, \citenamefont {McMahon},\ and\ \citenamefont {Byrnes}}]{Mohseni2022}%
  \BibitemOpen
  \bibfield  {author} {\bibinfo {author} {\bibfnamefont {N.}~\bibnamefont {Mohseni}}, \bibinfo {author} {\bibfnamefont {P.~L.}\ \bibnamefont {McMahon}},\ and\ \bibinfo {author} {\bibfnamefont {T.}~\bibnamefont {Byrnes}},\ }\bibfield  {title} {\bibinfo {title} {{Ising} machines as hardware solvers of combinatorial optimization problems},\ }\href {https://doi.org/10.1038/s42254-022-00440-8} {\bibfield  {journal} {\bibinfo  {journal} {Nat. Rev. Phys.}\ }\textbf {\bibinfo {volume} {4}},\ \bibinfo {pages} {363} (\bibinfo {year} {2022})}\BibitemShut {NoStop}%
\bibitem [{\citenamefont {Inagaki}\ \emph {et~al.}(2016)\citenamefont {Inagaki}, \citenamefont {Haribara}, \citenamefont {Igarashi}, \citenamefont {Sonobe}, \citenamefont {Tamate}, \citenamefont {Honjo}, \citenamefont {Marandi}, \citenamefont {McMahon}, \citenamefont {Umeki}, \citenamefont {Enbutsu}, \citenamefont {Tadanaga}, \citenamefont {Takenouchi}, \citenamefont {Aihara}, \citenamefont {Kawarabayashi}, \citenamefont {Inoue}, \citenamefont {Utsunomiya},\ and\ \citenamefont {Takesue}}]{Inagaki2016}%
  \BibitemOpen
  \bibfield  {author} {\bibinfo {author} {\bibfnamefont {T.}~\bibnamefont {Inagaki}}, \bibinfo {author} {\bibfnamefont {Y.}~\bibnamefont {Haribara}}, \bibinfo {author} {\bibfnamefont {K.}~\bibnamefont {Igarashi}}, \bibinfo {author} {\bibfnamefont {T.}~\bibnamefont {Sonobe}}, \bibinfo {author} {\bibfnamefont {S.}~\bibnamefont {Tamate}}, \bibinfo {author} {\bibfnamefont {T.}~\bibnamefont {Honjo}}, \bibinfo {author} {\bibfnamefont {A.}~\bibnamefont {Marandi}}, \bibinfo {author} {\bibfnamefont {P.~L.}\ \bibnamefont {McMahon}}, \bibinfo {author} {\bibfnamefont {T.}~\bibnamefont {Umeki}}, \bibinfo {author} {\bibfnamefont {K.}~\bibnamefont {Enbutsu}}, \bibinfo {author} {\bibfnamefont {O.}~\bibnamefont {Tadanaga}}, \bibinfo {author} {\bibfnamefont {H.}~\bibnamefont {Takenouchi}}, \bibinfo {author} {\bibfnamefont {K.}~\bibnamefont {Aihara}}, \bibinfo {author} {\bibfnamefont {K.-i.}\ \bibnamefont {Kawarabayashi}}, \bibinfo {author} {\bibfnamefont {K.}~\bibnamefont {Inoue}}, \bibinfo {author} {\bibfnamefont
  {S.}~\bibnamefont {Utsunomiya}},\ and\ \bibinfo {author} {\bibfnamefont {H.}~\bibnamefont {Takesue}},\ }\bibfield  {title} {\bibinfo {title} {A coherent {Ising} machine for 2000-node optimization problems},\ }\href {https://doi.org/10.1126/science.aah4243} {\bibfield  {journal} {\bibinfo  {journal} {Science}\ }\textbf {\bibinfo {volume} {354}},\ \bibinfo {pages} {603} (\bibinfo {year} {2016})}\BibitemShut {NoStop}%
\bibitem [{\citenamefont {McMahon}\ \emph {et~al.}(2016)\citenamefont {McMahon}, \citenamefont {Marandi}, \citenamefont {Haribara}, \citenamefont {Hamerly}, \citenamefont {Langrock}, \citenamefont {Tamate}, \citenamefont {Inagaki}, \citenamefont {Takesue}, \citenamefont {Utsunomiya}, \citenamefont {Aihara}, \citenamefont {Byer}, \citenamefont {Fejer}, \citenamefont {Mabuchi},\ and\ \citenamefont {Yamamoto}}]{McMahon2016}%
  \BibitemOpen
  \bibfield  {author} {\bibinfo {author} {\bibfnamefont {P.~L.}\ \bibnamefont {McMahon}}, \bibinfo {author} {\bibfnamefont {A.}~\bibnamefont {Marandi}}, \bibinfo {author} {\bibfnamefont {Y.}~\bibnamefont {Haribara}}, \bibinfo {author} {\bibfnamefont {R.}~\bibnamefont {Hamerly}}, \bibinfo {author} {\bibfnamefont {C.}~\bibnamefont {Langrock}}, \bibinfo {author} {\bibfnamefont {S.}~\bibnamefont {Tamate}}, \bibinfo {author} {\bibfnamefont {T.}~\bibnamefont {Inagaki}}, \bibinfo {author} {\bibfnamefont {H.}~\bibnamefont {Takesue}}, \bibinfo {author} {\bibfnamefont {S.}~\bibnamefont {Utsunomiya}}, \bibinfo {author} {\bibfnamefont {K.}~\bibnamefont {Aihara}}, \bibinfo {author} {\bibfnamefont {R.~L.}\ \bibnamefont {Byer}}, \bibinfo {author} {\bibfnamefont {M.~M.}\ \bibnamefont {Fejer}}, \bibinfo {author} {\bibfnamefont {H.}~\bibnamefont {Mabuchi}},\ and\ \bibinfo {author} {\bibfnamefont {Y.}~\bibnamefont {Yamamoto}},\ }\bibfield  {title} {\bibinfo {title} {A fully programmable 100-spin coherent {Ising} machine with
  all-to-all connections},\ }\href {https://doi.org/10.1126/science.aah5178} {\bibfield  {journal} {\bibinfo  {journal} {Science}\ }\textbf {\bibinfo {volume} {354}},\ \bibinfo {pages} {614} (\bibinfo {year} {2016})}\BibitemShut {NoStop}%
\bibitem [{\citenamefont {Yamamoto}\ \emph {et~al.}(2017)\citenamefont {Yamamoto}, \citenamefont {Aihara}, \citenamefont {Leleu}, \citenamefont {Kawarabayashi}, \citenamefont {Kako}, \citenamefont {Fejer}, \citenamefont {Inoue},\ and\ \citenamefont {Takesue}}]{Yamamoto2017}%
  \BibitemOpen
  \bibfield  {author} {\bibinfo {author} {\bibfnamefont {Y.}~\bibnamefont {Yamamoto}}, \bibinfo {author} {\bibfnamefont {K.}~\bibnamefont {Aihara}}, \bibinfo {author} {\bibfnamefont {T.}~\bibnamefont {Leleu}}, \bibinfo {author} {\bibfnamefont {K.-i.}\ \bibnamefont {Kawarabayashi}}, \bibinfo {author} {\bibfnamefont {S.}~\bibnamefont {Kako}}, \bibinfo {author} {\bibfnamefont {M.}~\bibnamefont {Fejer}}, \bibinfo {author} {\bibfnamefont {K.}~\bibnamefont {Inoue}},\ and\ \bibinfo {author} {\bibfnamefont {H.}~\bibnamefont {Takesue}},\ }\bibfield  {title} {\bibinfo {title} {Coherent {Ising} machines---optical neural networks operating at the quantum limit},\ }\href {https://doi.org/10.1038/s41534-017-0048-9} {\bibfield  {journal} {\bibinfo  {journal} {npj Quantum Inf.}\ }\textbf {\bibinfo {volume} {3}},\ \bibinfo {pages} {49} (\bibinfo {year} {2017})}\BibitemShut {NoStop}%
\bibitem [{\citenamefont {Honjo}\ \emph {et~al.}(2021)\citenamefont {Honjo}, \citenamefont {Sonobe}, \citenamefont {Inaba}, \citenamefont {Inagaki}, \citenamefont {Ikuta}, \citenamefont {Yamada}, \citenamefont {Kazama}, \citenamefont {Enbutsu}, \citenamefont {Umeki}, \citenamefont {Kasahara}, \citenamefont {Kawarabayashi},\ and\ \citenamefont {Takesue}}]{Honjo2021}%
  \BibitemOpen
  \bibfield  {author} {\bibinfo {author} {\bibfnamefont {T.}~\bibnamefont {Honjo}}, \bibinfo {author} {\bibfnamefont {T.}~\bibnamefont {Sonobe}}, \bibinfo {author} {\bibfnamefont {K.}~\bibnamefont {Inaba}}, \bibinfo {author} {\bibfnamefont {T.}~\bibnamefont {Inagaki}}, \bibinfo {author} {\bibfnamefont {T.}~\bibnamefont {Ikuta}}, \bibinfo {author} {\bibfnamefont {Y.}~\bibnamefont {Yamada}}, \bibinfo {author} {\bibfnamefont {T.}~\bibnamefont {Kazama}}, \bibinfo {author} {\bibfnamefont {K.}~\bibnamefont {Enbutsu}}, \bibinfo {author} {\bibfnamefont {T.}~\bibnamefont {Umeki}}, \bibinfo {author} {\bibfnamefont {R.}~\bibnamefont {Kasahara}}, \bibinfo {author} {\bibfnamefont {K.-i.}\ \bibnamefont {Kawarabayashi}},\ and\ \bibinfo {author} {\bibfnamefont {H.}~\bibnamefont {Takesue}},\ }\bibfield  {title} {\bibinfo {title} {100,000-spin coherent {Ising} machine},\ }\href {https://doi.org/10.1126/sciadv.abh0952} {\bibfield  {journal} {\bibinfo  {journal} {Sci. Adv.}\ }\textbf {\bibinfo {volume} {7}},\ \bibinfo {pages}
  {eabh0952} (\bibinfo {year} {2021})}\BibitemShut {NoStop}%
\bibitem [{\citenamefont {Rah}\ and\ \citenamefont {Yu}(2026)}]{Rah2026}%
  \BibitemOpen
  \bibfield  {author} {\bibinfo {author} {\bibfnamefont {Y.}~\bibnamefont {Rah}}\ and\ \bibinfo {author} {\bibfnamefont {K.}~\bibnamefont {Yu}},\ }\bibfield  {title} {\bibinfo {title} {Physical coherent {Ising} machines for solving combinatorial optimization problems},\ }\href {https://doi.org/10.35848/1347-4065/ae4034} {\bibfield  {journal} {\bibinfo  {journal} {Jpn. J. Appl. Phys.}\ }\textbf {\bibinfo {volume} {65}},\ \bibinfo {pages} {040803} (\bibinfo {year} {2026})}\BibitemShut {NoStop}%
\bibitem [{\citenamefont {Farhi}\ \emph {et~al.}(2014)\citenamefont {Farhi}, \citenamefont {Goldstone},\ and\ \citenamefont {Gutmann}}]{QAOA}%
  \BibitemOpen
  \bibfield  {author} {\bibinfo {author} {\bibfnamefont {E.}~\bibnamefont {Farhi}}, \bibinfo {author} {\bibfnamefont {J.}~\bibnamefont {Goldstone}},\ and\ \bibinfo {author} {\bibfnamefont {S.}~\bibnamefont {Gutmann}},\ }\href {https://doi.org/10.48550/arXiv.1411.4028} {\bibinfo {title} {A quantum approximate optimization algorithm}} (\bibinfo {year} {2014}),\ \Eprint {https://arxiv.org/abs/1411.4028} {arXiv:1411.4028 [quant-ph]} \BibitemShut {NoStop}%
\bibitem [{\citenamefont {Hadfield}\ \emph {et~al.}(2019)\citenamefont {Hadfield}, \citenamefont {Wang}, \citenamefont {O'Gorman}, \citenamefont {Rieffel}, \citenamefont {Venturelli},\ and\ \citenamefont {Biswas}}]{Hadfield2019}%
  \BibitemOpen
  \bibfield  {author} {\bibinfo {author} {\bibfnamefont {S.}~\bibnamefont {Hadfield}}, \bibinfo {author} {\bibfnamefont {Z.}~\bibnamefont {Wang}}, \bibinfo {author} {\bibfnamefont {B.}~\bibnamefont {O'Gorman}}, \bibinfo {author} {\bibfnamefont {E.~G.}\ \bibnamefont {Rieffel}}, \bibinfo {author} {\bibfnamefont {D.}~\bibnamefont {Venturelli}},\ and\ \bibinfo {author} {\bibfnamefont {R.}~\bibnamefont {Biswas}},\ }\bibfield  {title} {\bibinfo {title} {From the quantum approximate optimization algorithm to a quantum alternating operator ansatz},\ }\href {https://doi.org/10.3390/a12020034} {\bibfield  {journal} {\bibinfo  {journal} {Algorithms}\ }\textbf {\bibinfo {volume} {12}},\ \bibinfo {pages} {34} (\bibinfo {year} {2019})}\BibitemShut {NoStop}%
\bibitem [{\citenamefont {Cerezo}\ \emph {et~al.}(2021)\citenamefont {Cerezo}, \citenamefont {Arrasmith}, \citenamefont {Babbush}, \citenamefont {Benjamin}, \citenamefont {Endo}, \citenamefont {Fujii}, \citenamefont {McClean}, \citenamefont {Mitarai}, \citenamefont {Yuan}, \citenamefont {Cincio},\ and\ \citenamefont {Coles}}]{Cerezo2021}%
  \BibitemOpen
  \bibfield  {author} {\bibinfo {author} {\bibfnamefont {M.}~\bibnamefont {Cerezo}}, \bibinfo {author} {\bibfnamefont {A.}~\bibnamefont {Arrasmith}}, \bibinfo {author} {\bibfnamefont {R.}~\bibnamefont {Babbush}}, \bibinfo {author} {\bibfnamefont {S.~C.}\ \bibnamefont {Benjamin}}, \bibinfo {author} {\bibfnamefont {S.}~\bibnamefont {Endo}}, \bibinfo {author} {\bibfnamefont {K.}~\bibnamefont {Fujii}}, \bibinfo {author} {\bibfnamefont {J.~R.}\ \bibnamefont {McClean}}, \bibinfo {author} {\bibfnamefont {K.}~\bibnamefont {Mitarai}}, \bibinfo {author} {\bibfnamefont {X.}~\bibnamefont {Yuan}}, \bibinfo {author} {\bibfnamefont {L.}~\bibnamefont {Cincio}},\ and\ \bibinfo {author} {\bibfnamefont {P.~J.}\ \bibnamefont {Coles}},\ }\bibfield  {title} {\bibinfo {title} {Variational quantum algorithms},\ }\href {https://doi.org/10.1038/s42254-021-00348-9} {\bibfield  {journal} {\bibinfo  {journal} {Nat. Rev. Phys.}\ }\textbf {\bibinfo {volume} {3}},\ \bibinfo {pages} {625} (\bibinfo {year} {2021})}\BibitemShut {NoStop}%
\bibitem [{\citenamefont {Tilly}\ \emph {et~al.}(2022)\citenamefont {Tilly}, \citenamefont {Chen}, \citenamefont {Cao}, \citenamefont {Picozzi}, \citenamefont {Setia}, \citenamefont {Li}, \citenamefont {Grant}, \citenamefont {Wossnig}, \citenamefont {Rungger}, \citenamefont {Booth},\ and\ \citenamefont {Tennyson}}]{Tilly2022}%
  \BibitemOpen
  \bibfield  {author} {\bibinfo {author} {\bibfnamefont {J.}~\bibnamefont {Tilly}}, \bibinfo {author} {\bibfnamefont {H.}~\bibnamefont {Chen}}, \bibinfo {author} {\bibfnamefont {S.}~\bibnamefont {Cao}}, \bibinfo {author} {\bibfnamefont {D.}~\bibnamefont {Picozzi}}, \bibinfo {author} {\bibfnamefont {K.}~\bibnamefont {Setia}}, \bibinfo {author} {\bibfnamefont {Y.}~\bibnamefont {Li}}, \bibinfo {author} {\bibfnamefont {E.}~\bibnamefont {Grant}}, \bibinfo {author} {\bibfnamefont {L.}~\bibnamefont {Wossnig}}, \bibinfo {author} {\bibfnamefont {I.}~\bibnamefont {Rungger}}, \bibinfo {author} {\bibfnamefont {G.~H.}\ \bibnamefont {Booth}},\ and\ \bibinfo {author} {\bibfnamefont {J.}~\bibnamefont {Tennyson}},\ }\bibfield  {title} {\bibinfo {title} {The variational quantum eigensolver: a review of methods and best practices},\ }\href {https://doi.org/10.1016/j.physrep.2022.08.003} {\bibfield  {journal} {\bibinfo  {journal} {Phys. Rep.}\ }\textbf {\bibinfo {volume} {986}},\ \bibinfo {pages} {1} (\bibinfo {year}
  {2022})}\BibitemShut {NoStop}%
\bibitem [{\citenamefont {Kiktenko}\ \emph {et~al.}(2025)\citenamefont {Kiktenko}, \citenamefont {Krendeleva},\ and\ \citenamefont {Fedorov}}]{Kiktenko2025}%
  \BibitemOpen
  \bibfield  {author} {\bibinfo {author} {\bibfnamefont {E.~O.}\ \bibnamefont {Kiktenko}}, \bibinfo {author} {\bibfnamefont {E.~V.}\ \bibnamefont {Krendeleva}},\ and\ \bibinfo {author} {\bibfnamefont {A.~K.}\ \bibnamefont {Fedorov}},\ }\href {https://doi.org/10.48550/arXiv.2512.23026} {\bibinfo {title} {Applying grover-mixer quantum alternating operator ansatz algorithm to high-order unconstrained binary optimization problems}} (\bibinfo {year} {2025}),\ \Eprint {https://arxiv.org/abs/2512.23026} {arXiv:2512.23026 [quant-ph]} \BibitemShut {NoStop}%
\bibitem [{\citenamefont {Leleu}\ \emph {et~al.}(2019)\citenamefont {Leleu}, \citenamefont {Yamamoto}, \citenamefont {McMahon},\ and\ \citenamefont {Aihara}}]{Leleu2019}%
  \BibitemOpen
  \bibfield  {author} {\bibinfo {author} {\bibfnamefont {T.}~\bibnamefont {Leleu}}, \bibinfo {author} {\bibfnamefont {Y.}~\bibnamefont {Yamamoto}}, \bibinfo {author} {\bibfnamefont {P.~L.}\ \bibnamefont {McMahon}},\ and\ \bibinfo {author} {\bibfnamefont {K.}~\bibnamefont {Aihara}},\ }\bibfield  {title} {\bibinfo {title} {Destabilization of local minima in analog spin systems by correction of amplitude heterogeneity},\ }\href {https://doi.org/10.1103/PhysRevLett.122.040607} {\bibfield  {journal} {\bibinfo  {journal} {Phys. Rev. Lett.}\ }\textbf {\bibinfo {volume} {122}},\ \bibinfo {pages} {040607} (\bibinfo {year} {2019})}\BibitemShut {NoStop}%
\bibitem [{\citenamefont {Hamerly}\ \emph {et~al.}(2019)\citenamefont {Hamerly}, \citenamefont {Inagaki}, \citenamefont {McMahon}, \citenamefont {Venturelli}, \citenamefont {Marandi}, \citenamefont {Onodera}, \citenamefont {Ng}, \citenamefont {Langrock}, \citenamefont {Inaba}, \citenamefont {Honjo}, \citenamefont {Enbutsu}, \citenamefont {Umeki}, \citenamefont {Kasahara}, \citenamefont {Utsunomiya}, \citenamefont {Kako}, \citenamefont {Kawarabayashi}, \citenamefont {Byer}, \citenamefont {Fejer}, \citenamefont {Mabuchi}, \citenamefont {Englund}, \citenamefont {Rieffel}, \citenamefont {Takesue},\ and\ \citenamefont {Yamamoto}}]{Hamerly2019}%
  \BibitemOpen
  \bibfield  {author} {\bibinfo {author} {\bibfnamefont {R.}~\bibnamefont {Hamerly}}, \bibinfo {author} {\bibfnamefont {T.}~\bibnamefont {Inagaki}}, \bibinfo {author} {\bibfnamefont {P.~L.}\ \bibnamefont {McMahon}}, \bibinfo {author} {\bibfnamefont {D.}~\bibnamefont {Venturelli}}, \bibinfo {author} {\bibfnamefont {A.}~\bibnamefont {Marandi}}, \bibinfo {author} {\bibfnamefont {T.}~\bibnamefont {Onodera}}, \bibinfo {author} {\bibfnamefont {E.}~\bibnamefont {Ng}}, \bibinfo {author} {\bibfnamefont {C.}~\bibnamefont {Langrock}}, \bibinfo {author} {\bibfnamefont {K.}~\bibnamefont {Inaba}}, \bibinfo {author} {\bibfnamefont {T.}~\bibnamefont {Honjo}}, \bibinfo {author} {\bibfnamefont {K.}~\bibnamefont {Enbutsu}}, \bibinfo {author} {\bibfnamefont {T.}~\bibnamefont {Umeki}}, \bibinfo {author} {\bibfnamefont {R.}~\bibnamefont {Kasahara}}, \bibinfo {author} {\bibfnamefont {S.}~\bibnamefont {Utsunomiya}}, \bibinfo {author} {\bibfnamefont {S.}~\bibnamefont {Kako}}, \bibinfo {author} {\bibfnamefont {K.-i.}\ \bibnamefont
  {Kawarabayashi}}, \bibinfo {author} {\bibfnamefont {R.~L.}\ \bibnamefont {Byer}}, \bibinfo {author} {\bibfnamefont {M.~M.}\ \bibnamefont {Fejer}}, \bibinfo {author} {\bibfnamefont {H.}~\bibnamefont {Mabuchi}}, \bibinfo {author} {\bibfnamefont {D.}~\bibnamefont {Englund}}, \bibinfo {author} {\bibfnamefont {E.}~\bibnamefont {Rieffel}}, \bibinfo {author} {\bibfnamefont {H.}~\bibnamefont {Takesue}},\ and\ \bibinfo {author} {\bibfnamefont {Y.}~\bibnamefont {Yamamoto}},\ }\bibfield  {title} {\bibinfo {title} {Experimental investigation of performance differences between coherent {Ising} machines and a quantum annealer},\ }\href {https://doi.org/10.1126/sciadv.aau0823} {\bibfield  {journal} {\bibinfo  {journal} {Sci. Adv.}\ }\textbf {\bibinfo {volume} {5}},\ \bibinfo {pages} {eaau0823} (\bibinfo {year} {2019})}\BibitemShut {NoStop}%
\bibitem [{\citenamefont {Dobrynin}\ \emph {et~al.}(2024)\citenamefont {Dobrynin}, \citenamefont {Renaudineau}, \citenamefont {Hizzani}, \citenamefont {Strukov}, \citenamefont {Mohseni},\ and\ \citenamefont {Strachan}}]{Dobrynin2024}%
  \BibitemOpen
  \bibfield  {author} {\bibinfo {author} {\bibfnamefont {D.~A.}\ \bibnamefont {Dobrynin}}, \bibinfo {author} {\bibfnamefont {A.}~\bibnamefont {Renaudineau}}, \bibinfo {author} {\bibfnamefont {M.}~\bibnamefont {Hizzani}}, \bibinfo {author} {\bibfnamefont {D.}~\bibnamefont {Strukov}}, \bibinfo {author} {\bibfnamefont {M.}~\bibnamefont {Mohseni}},\ and\ \bibinfo {author} {\bibfnamefont {J.~P.}\ \bibnamefont {Strachan}},\ }\bibfield  {title} {\bibinfo {title} {Energy landscapes of combinatorial optimization in {Ising} machines},\ }\href {https://doi.org/10.1103/PhysRevE.110.045308} {\bibfield  {journal} {\bibinfo  {journal} {Phys. Rev. E}\ }\textbf {\bibinfo {volume} {110}},\ \bibinfo {pages} {045308} (\bibinfo {year} {2024})}\BibitemShut {NoStop}%
\bibitem [{\citenamefont {Zhou}\ \emph {et~al.}(2025)\citenamefont {Zhou}, \citenamefont {Wong}, \citenamefont {Wang}, \citenamefont {Hui}, \citenamefont {Ebler},\ and\ \citenamefont {Sun}}]{Zhou2025}%
  \BibitemOpen
  \bibfield  {author} {\bibinfo {author} {\bibfnamefont {S.}~\bibnamefont {Zhou}}, \bibinfo {author} {\bibfnamefont {K.~Y.~M.}\ \bibnamefont {Wong}}, \bibinfo {author} {\bibfnamefont {J.}~\bibnamefont {Wang}}, \bibinfo {author} {\bibfnamefont {D.~S.~W.}\ \bibnamefont {Hui}}, \bibinfo {author} {\bibfnamefont {D.}~\bibnamefont {Ebler}},\ and\ \bibinfo {author} {\bibfnamefont {J.}~\bibnamefont {Sun}},\ }\href {https://doi.org/10.48550/arXiv.2507.08533} {\bibinfo {title} {Phase analysis of {Ising} machines and their implications on optimization}} (\bibinfo {year} {2025}),\ \Eprint {https://arxiv.org/abs/2507.08533} {arXiv:2507.08533 [cond-mat.dis-nn]} \BibitemShut {NoStop}%
\bibitem [{\citenamefont {Grover}(1996)}]{Grover1996}%
  \BibitemOpen
  \bibfield  {author} {\bibinfo {author} {\bibfnamefont {L.~K.}\ \bibnamefont {Grover}},\ }\bibfield  {title} {\bibinfo {title} {A fast quantum mechanical algorithm for database search},\ }in\ \href {https://doi.org/10.1145/237814.237866} {\emph {\bibinfo {booktitle} {Proceedings of the Twenty-Eighth Annual {ACM} Symposium on Theory of Computing}}},\ \bibinfo {series and number} {{STOC} '96}\ (\bibinfo  {publisher} {Association for Computing Machinery},\ \bibinfo {address} {New York, NY, USA},\ \bibinfo {year} {1996})\ pp.\ \bibinfo {pages} {212--219}\BibitemShut {NoStop}%
\bibitem [{\citenamefont {Bennett}\ \emph {et~al.}(1997)\citenamefont {Bennett}, \citenamefont {Bernstein}, \citenamefont {Brassard},\ and\ \citenamefont {Vazirani}}]{Bennett1997}%
  \BibitemOpen
  \bibfield  {author} {\bibinfo {author} {\bibfnamefont {C.~H.}\ \bibnamefont {Bennett}}, \bibinfo {author} {\bibfnamefont {E.}~\bibnamefont {Bernstein}}, \bibinfo {author} {\bibfnamefont {G.}~\bibnamefont {Brassard}},\ and\ \bibinfo {author} {\bibfnamefont {U.}~\bibnamefont {Vazirani}},\ }\bibfield  {title} {\bibinfo {title} {Strengths and weaknesses of quantum computing},\ }\href {https://doi.org/10.1137/S0097539796300933} {\bibfield  {journal} {\bibinfo  {journal} {SIAM J. Comput.}\ }\textbf {\bibinfo {volume} {26}},\ \bibinfo {pages} {1510} (\bibinfo {year} {1997})}\BibitemShut {NoStop}%
\bibitem [{\citenamefont {Boyer}\ \emph {et~al.}(1998)\citenamefont {Boyer}, \citenamefont {Brassard}, \citenamefont {H{\o}yer},\ and\ \citenamefont {Tapp}}]{Boyer1998}%
  \BibitemOpen
  \bibfield  {author} {\bibinfo {author} {\bibfnamefont {M.}~\bibnamefont {Boyer}}, \bibinfo {author} {\bibfnamefont {G.}~\bibnamefont {Brassard}}, \bibinfo {author} {\bibfnamefont {P.}~\bibnamefont {H{\o}yer}},\ and\ \bibinfo {author} {\bibfnamefont {A.}~\bibnamefont {Tapp}},\ }\bibfield  {title} {\bibinfo {title} {Tight bounds on quantum searching},\ }\href {https://doi.org/10.1002/(SICI)1521-3978(199806)46:4/5<493::AID-PROP493>3.0.CO;2-P} {\bibfield  {journal} {\bibinfo  {journal} {Fortschr. Phys.}\ }\textbf {\bibinfo {volume} {46}},\ \bibinfo {pages} {493} (\bibinfo {year} {1998})}\BibitemShut {NoStop}%
\bibitem [{\citenamefont {Brassard}\ \emph {et~al.}(2002)\citenamefont {Brassard}, \citenamefont {H{\o}yer}, \citenamefont {Mosca},\ and\ \citenamefont {Tapp}}]{Brassard2002}%
  \BibitemOpen
  \bibfield  {author} {\bibinfo {author} {\bibfnamefont {G.}~\bibnamefont {Brassard}}, \bibinfo {author} {\bibfnamefont {P.}~\bibnamefont {H{\o}yer}}, \bibinfo {author} {\bibfnamefont {M.}~\bibnamefont {Mosca}},\ and\ \bibinfo {author} {\bibfnamefont {A.}~\bibnamefont {Tapp}},\ }\bibfield  {title} {\bibinfo {title} {Quantum amplitude amplification and estimation},\ }in\ \href {https://doi.org/10.1090/conm/305/05215} {\emph {\bibinfo {booktitle} {Quantum Computation and Information}}},\ \bibinfo {series} {Contemporary Mathematics}, Vol.\ \bibinfo {volume} {305},\ \bibinfo {editor} {edited by\ \bibinfo {editor} {\bibfnamefont {S.~J.}\ \bibnamefont {Lomonaco}}}\ (\bibinfo  {publisher} {American Mathematical Society},\ \bibinfo {address} {Providence, RI},\ \bibinfo {year} {2002})\ pp.\ \bibinfo {pages} {53--74}\BibitemShut {NoStop}%
\bibitem [{\citenamefont {Nielsen}\ and\ \citenamefont {Chuang}(2010)}]{Nielsen2010}%
  \BibitemOpen
  \bibfield  {author} {\bibinfo {author} {\bibfnamefont {M.~A.}\ \bibnamefont {Nielsen}}\ and\ \bibinfo {author} {\bibfnamefont {I.~L.}\ \bibnamefont {Chuang}},\ }\href {https://doi.org/10.1017/CBO9780511976667} {\emph {\bibinfo {title} {Quantum Computation and Quantum Information}}},\ \bibinfo {edition} {10th}\ ed.\ (\bibinfo  {publisher} {Cambridge University Press},\ \bibinfo {address} {Cambridge},\ \bibinfo {year} {2010})\BibitemShut {NoStop}%
\bibitem [{\citenamefont {D{\"u}rr}\ and\ \citenamefont {H{\o}yer}(1996)}]{DurrHoyer1996}%
  \BibitemOpen
  \bibfield  {author} {\bibinfo {author} {\bibfnamefont {C.}~\bibnamefont {D{\"u}rr}}\ and\ \bibinfo {author} {\bibfnamefont {P.}~\bibnamefont {H{\o}yer}},\ }\href {https://doi.org/10.48550/arXiv.quant-ph/9607014} {\bibinfo {title} {A quantum algorithm for finding the minimum}} (\bibinfo {year} {1996}),\ \Eprint {https://arxiv.org/abs/quant-ph/9607014} {arXiv:quant-ph/9607014} \BibitemShut {NoStop}%
\bibitem{Gilliam2021}
A. Gilliam, S. Woerner, and C. Gonciulea,
\href{https://doi.org/10.22331/q-2021-04-08-428}{Grover Adaptive Search for constrained polynomial binary optimization, Quantum \textbf{5}, 428 (2021)}.
\bibitem{LinTong361}
L. Lin and Y. Tong,
\href{https://doi.org/10.22331/q-2020-11-11-361}{Optimal polynomial based quantum eigenstate filtering with application to solving quantum linear systems, Quantum \textbf{4}, 361 (2020)}.
\bibitem{LinTong372}
L. Lin and Y. Tong,
\href{https://doi.org/10.22331/q-2020-12-14-372}{Near-optimal ground state preparation, Quantum \textbf{4}, 372 (2020)}.
\bibitem{Dong2022}
Y. Dong, L. Lin, and Y. Tong,
\href{https://doi.org/10.1103/PRXQuantum.3.040305}{Ground-state preparation and energy estimation on early fault-tolerant quantum computers via quantum eigenvalue transformation of unitary matrices, PRX Quantum \textbf{3}, 040305 (2022)}.
\bibitem{GeTuraCirac}
Y. Ge, J. Tura, and J. I. Cirac,
\href{https://doi.org/10.1063/1.5027484}{Faster ground state preparation and high-precision ground energy estimation with fewer qubits, J. Math. Phys. \textbf{60}, 022202 (2019)}.
\bibitem [{\citenamefont {Farhi}\ and\ \citenamefont {Gutmann}(1998)}]{FarhiGutmann1998}%
  \BibitemOpen
  \bibfield  {author} {\bibinfo {author} {\bibfnamefont {E.}~\bibnamefont {Farhi}}\ and\ \bibinfo {author} {\bibfnamefont {S.}~\bibnamefont {Gutmann}},\ }\bibfield  {title} {\bibinfo {title} {An analog analogue of a digital quantum computation},\ }\href {https://doi.org/10.1103/PhysRevA.57.2403} {\bibfield  {journal} {\bibinfo  {journal} {Phys. Rev. A}\ }\textbf {\bibinfo {volume} {57}},\ \bibinfo {pages} {2403} (\bibinfo {year} {1998})}\BibitemShut {NoStop}%
\bibitem [{\citenamefont {Jiang}\ \emph {et~al.}(2017)\citenamefont {Jiang}, \citenamefont {Rieffel},\ and\ \citenamefont {Wang}}]{Jiang2017}%
  \BibitemOpen
  \bibfield  {author} {\bibinfo {author} {\bibfnamefont {Z.}~\bibnamefont {Jiang}}, \bibinfo {author} {\bibfnamefont {E.~G.}\ \bibnamefont {Rieffel}},\ and\ \bibinfo {author} {\bibfnamefont {Z.}~\bibnamefont {Wang}},\ }\bibfield  {title} {\bibinfo {title} {Near-optimal quantum circuit for {Grover}'s unstructured search using a transverse field},\ }\href {https://doi.org/10.1103/PhysRevA.95.062317} {\bibfield  {journal} {\bibinfo  {journal} {Phys. Rev. A}\ }\textbf {\bibinfo {volume} {95}},\ \bibinfo {pages} {062317} (\bibinfo {year} {2017})}\BibitemShut {NoStop}%
\bibitem [{\citenamefont {H{\o}yer}(2000)}]{Hoyer2000}%
  \BibitemOpen
  \bibfield  {author} {\bibinfo {author} {\bibfnamefont {P.}~\bibnamefont {H{\o}yer}},\ }\bibfield  {title} {\bibinfo {title} {Arbitrary phases in quantum amplitude amplification},\ }\href {https://doi.org/10.1103/PhysRevA.62.052304} {\bibfield  {journal} {\bibinfo  {journal} {Phys. Rev. A}\ }\textbf {\bibinfo {volume} {62}},\ \bibinfo {pages} {052304} (\bibinfo {year} {2000})}\BibitemShut {NoStop}%
\bibitem [{\citenamefont {Yoder}\ \emph {et~al.}(2014)\citenamefont {Yoder}, \citenamefont {Low},\ and\ \citenamefont {Chuang}}]{Yoder2014}%
  \BibitemOpen
  \bibfield  {author} {\bibinfo {author} {\bibfnamefont {T.~J.}\ \bibnamefont {Yoder}}, \bibinfo {author} {\bibfnamefont {G.~H.}\ \bibnamefont {Low}},\ and\ \bibinfo {author} {\bibfnamefont {I.~L.}\ \bibnamefont {Chuang}},\ }\bibfield  {title} {\bibinfo {title} {Fixed-point quantum search with an optimal number of queries},\ }\href {https://doi.org/10.1103/PhysRevLett.113.210501} {\bibfield  {journal} {\bibinfo  {journal} {Phys. Rev. Lett.}\ }\textbf {\bibinfo {volume} {113}},\ \bibinfo {pages} {210501} (\bibinfo {year} {2014})}\BibitemShut {NoStop}%
\bibitem [{\citenamefont {Anikeeva}\ \emph {et~al.}(2021)\citenamefont {Anikeeva}, \citenamefont {Markovi{\'c}}, \citenamefont {Borish}, \citenamefont {Hines}, \citenamefont {Rajagopal}, \citenamefont {Cooper}, \citenamefont {Periwal}, \citenamefont {Safavi-Naeini}, \citenamefont {Davis},\ and\ \citenamefont {Schleier-Smith}}]{Anikeeva2021}%
  \BibitemOpen
  \bibfield  {author} {\bibinfo {author} {\bibfnamefont {G.}~\bibnamefont {Anikeeva}}, \bibinfo {author} {\bibfnamefont {O.}~\bibnamefont {Markovi{\'c}}}, \bibinfo {author} {\bibfnamefont {V.}~\bibnamefont {Borish}}, \bibinfo {author} {\bibfnamefont {J.~A.}\ \bibnamefont {Hines}}, \bibinfo {author} {\bibfnamefont {S.~V.}\ \bibnamefont {Rajagopal}}, \bibinfo {author} {\bibfnamefont {E.~S.}\ \bibnamefont {Cooper}}, \bibinfo {author} {\bibfnamefont {A.}~\bibnamefont {Periwal}}, \bibinfo {author} {\bibfnamefont {A.}~\bibnamefont {Safavi-Naeini}}, \bibinfo {author} {\bibfnamefont {E.~J.}\ \bibnamefont {Davis}},\ and\ \bibinfo {author} {\bibfnamefont {M.}~\bibnamefont {Schleier-Smith}},\ }\bibfield  {title} {\bibinfo {title} {Number partitioning with {Grover}'s algorithm in central spin systems},\ }\href {https://doi.org/10.1103/PRXQuantum.2.020319} {\bibfield  {journal} {\bibinfo  {journal} {PRX Quantum}\ }\textbf {\bibinfo {volume} {2}},\ \bibinfo {pages} {020319} (\bibinfo {year} {2021})}\BibitemShut {NoStop}%
\bibitem [{\citenamefont {Sinitsyn}\ and\ \citenamefont {Yan}(2023)}]{Sinitsyn2023}%
  \BibitemOpen
  \bibfield  {author} {\bibinfo {author} {\bibfnamefont {N.~A.}\ \bibnamefont {Sinitsyn}}\ and\ \bibinfo {author} {\bibfnamefont {B.}~\bibnamefont {Yan}},\ }\bibfield  {title} {\bibinfo {title} {Topologically protected {Grover}'s oracle for the partition problem},\ }\href {https://doi.org/10.1103/PhysRevA.108.022412} {\bibfield  {journal} {\bibinfo  {journal} {Phys. Rev. A}\ }\textbf {\bibinfo {volume} {108}},\ \bibinfo {pages} {022412} (\bibinfo {year} {2023})}\BibitemShut {NoStop}%
\bibitem [{\citenamefont {Nieman}\ \emph {et~al.}(2024)\citenamefont {Nieman}, \citenamefont {Durand}, \citenamefont {Patel}, \citenamefont {Koch},\ and\ \citenamefont {Alsing}}]{Nieman2024}%
  \BibitemOpen
  \bibfield  {author} {\bibinfo {author} {\bibfnamefont {K.}~\bibnamefont {Nieman}}, \bibinfo {author} {\bibfnamefont {H.}~\bibnamefont {Durand}}, \bibinfo {author} {\bibfnamefont {S.}~\bibnamefont {Patel}}, \bibinfo {author} {\bibfnamefont {D.}~\bibnamefont {Koch}},\ and\ \bibinfo {author} {\bibfnamefont {P.~M.}\ \bibnamefont {Alsing}},\ }\bibfield  {title} {\bibinfo {title} {Investigating an amplitude amplification-based optimization algorithm for model predictive control},\ }\href {https://doi.org/10.1016/j.dche.2023.100134} {\bibfield  {journal} {\bibinfo  {journal} {Digit. Chem. Eng.}\ }\textbf {\bibinfo {volume} {10}},\ \bibinfo {pages} {100134} (\bibinfo {year} {2024})}\BibitemShut {NoStop}%
\bibitem [{\citenamefont {Koch}\ \emph {et~al.}(2022)\citenamefont {Koch}, \citenamefont {Cutugno}, \citenamefont {Karlson}, \citenamefont {Patel}, \citenamefont {Wessing},\ and\ \citenamefont {Alsing}}]{Koch2022}%
  \BibitemOpen
  \bibfield  {author} {\bibinfo {author} {\bibfnamefont {D.}~\bibnamefont {Koch}}, \bibinfo {author} {\bibfnamefont {M.}~\bibnamefont {Cutugno}}, \bibinfo {author} {\bibfnamefont {S.}~\bibnamefont {Karlson}}, \bibinfo {author} {\bibfnamefont {S.}~\bibnamefont {Patel}}, \bibinfo {author} {\bibfnamefont {L.}~\bibnamefont {Wessing}},\ and\ \bibinfo {author} {\bibfnamefont {P.~M.}\ \bibnamefont {Alsing}},\ }\bibfield  {title} {\bibinfo {title} {Gaussian amplitude amplification for quantum pathfinding},\ }\href {https://doi.org/10.3390/e24070963} {\bibfield  {journal} {\bibinfo  {journal} {Entropy}\ }\textbf {\bibinfo {volume} {24}},\ \bibinfo {pages} {963} (\bibinfo {year} {2022})}\BibitemShut {NoStop}%
\bibitem [{\citenamefont {Yan}\ \emph {et~al.}(2022)\citenamefont {Yan}, \citenamefont {Wei}, \citenamefont {Jiang}, \citenamefont {Wang}, \citenamefont {Duan}, \citenamefont {Ma},\ and\ \citenamefont {Long}}]{Yan2022}%
  \BibitemOpen
  \bibfield  {author} {\bibinfo {author} {\bibfnamefont {B.}~\bibnamefont {Yan}}, \bibinfo {author} {\bibfnamefont {S.}~\bibnamefont {Wei}}, \bibinfo {author} {\bibfnamefont {H.}~\bibnamefont {Jiang}}, \bibinfo {author} {\bibfnamefont {H.}~\bibnamefont {Wang}}, \bibinfo {author} {\bibfnamefont {Q.}~\bibnamefont {Duan}}, \bibinfo {author} {\bibfnamefont {Z.}~\bibnamefont {Ma}},\ and\ \bibinfo {author} {\bibfnamefont {G.-L.}\ \bibnamefont {Long}},\ }\bibfield  {title} {\bibinfo {title} {Fixed-point oblivious quantum amplitude-amplification algorithm},\ }\href {https://doi.org/10.1038/s41598-022-15093-x} {\bibfield  {journal} {\bibinfo  {journal} {Sci. Rep.}\ }\textbf {\bibinfo {volume} {12}},\ \bibinfo {pages} {14339} (\bibinfo {year} {2022})}\BibitemShut {NoStop}%
\bibitem [{\citenamefont {Zhukov}\ \emph {et~al.}(2025)\citenamefont {Zhukov}, \citenamefont {Lebedev},\ and\ \citenamefont {Pogosov}}]{PreviousPaper}%
  \BibitemOpen
  \bibfield  {author} {\bibinfo {author} {\bibfnamefont {A.~A.}\ \bibnamefont {Zhukov}}, \bibinfo {author} {\bibfnamefont {A.~V.}\ \bibnamefont {Lebedev}},\ and\ \bibinfo {author} {\bibfnamefont {W.~V.}\ \bibnamefont {Pogosov}},\ }\bibfield  {title} {\bibinfo {title} {{Grover}'s search meets {Ising} models: A quantum algorithm for finding low-energy states},\ }\href {https://doi.org/10.1016/j.cpc.2025.109627} {\bibfield  {journal} {\bibinfo  {journal} {Comput. Phys. Commun.}\ }\textbf {\bibinfo {volume} {313}},\ \bibinfo {pages} {109627} (\bibinfo {year} {2025})}\BibitemShut {NoStop}%
\bibitem{Talkner2007}
P. Talkner, E. Lutz, and P. H{\"a}nggi,
\href{https://doi.org/10.1103/PhysRevE.75.050102}{Fluctuation theorems: Work is not an observable, Phys. Rev. E \textbf{75}, 050102(R) (2007)}.
\bibitem{Heyl2013}
M. Heyl, A. Polkovnikov, and S. Kehrein,
\href{https://doi.org/10.1103/PhysRevLett.110.135704}{Dynamical quantum phase transitions in the transverse-field Ising model, Phys. Rev. Lett. \textbf{110}, 135704 (2013)}.
\bibitem [{\citenamefont {Mezard}\ \emph {et~al.}(1987)\citenamefont {Mezard}, \citenamefont {Parisi},\ and\ \citenamefont {Virasoro}}]{Parisi}%
  \BibitemOpen
  \bibfield  {author} {\bibinfo {author} {\bibfnamefont {M.}~\bibnamefont {Mezard}}, \bibinfo {author} {\bibfnamefont {G.}~\bibnamefont {Parisi}},\ and\ \bibinfo {author} {\bibfnamefont {M.~A.}\ \bibnamefont {Virasoro}},\ }\href@noop {} {\emph {\bibinfo {title} {Spin Glass Theory and Beyond}}}\ (\bibinfo  {publisher} {World Scientific},\ \bibinfo {address} {Singapore},\ \bibinfo {year} {1987})\BibitemShut {NoStop}%
\bibitem{Shenvi2003}
N. Shenvi, K. R. Brown, and K. B. Whaley,
\href{https://doi.org/10.1103/PhysRevA.68.052313}{Effects of a random noisy oracle on search algorithm complexity, Phys. Rev. A \textbf{68}, 052313 (2003)}.
\bibitem{RossSelinger2016}
N. J. Ross and P. Selinger,
\href{https://doi.org/10.26421/QIC16.11-12-1}{Optimal ancilla-free Clifford+$T$ approximation of $z$-rotations, Quantum Inf. Comput. \textbf{16}, 901 (2016)}.
\bibitem [{\citenamefont {Plyashechnik}\ \emph {et~al.}(2026)\citenamefont {Plyashechnik}, \citenamefont {Zhukov}, \citenamefont {Lebedev},\ and\ \citenamefont {Pogosov}}]{Phi0ZenodoData}%
  \BibitemOpen
  \bibfield  {author} {\bibinfo {author} {\bibfnamefont {A.~S.}\ \bibnamefont {Plyashechnik}}, \bibinfo {author} {\bibfnamefont {A.~A.}\ \bibnamefont {Zhukov}}, \bibinfo {author} {\bibfnamefont {A.~V.}\ \bibnamefont {Lebedev}},\ and\ \bibinfo {author} {\bibfnamefont {W.~V.}\ \bibnamefont {Pogosov}},\ }\href {https://doi.org/10.5281/zenodo.20376132} {\bibinfo {title} {Data and code for ``energy-selective quantum search with ising hamiltonian phase oracles''}} (\bibinfo {year} {2026})\BibitemShut {NoStop}%
\end{thebibliography}
\end{document}